\documentclass{aa}  

\usepackage{graphicx}
\usepackage{txfonts}
\usepackage{lipsum}
\usepackage{subcaption}         
\usepackage{lscape}             
\usepackage{placeins}           
                                
\usepackage[bookmarks]{hyperref}
\usepackage{natbib}
\usepackage[T1]{fontenc}
\usepackage{amsmath}
\usepackage{lmodern}
\usepackage{amssymb}
\usepackage{changepage}
\usepackage{booktabs}
\usepackage{rotating}
\usepackage{caption}
\bibpunct{(}{)}{;}{a}{}{,} 
\usepackage{tabularx}
\usepackage[table]{xcolor}
\usepackage{caption}
\usepackage{longtable}
\usepackage{xltabular}
\usepackage{array}

\hypersetup{
    colorlinks=true,       
    linkcolor=blue,       
    citecolor=blue,       
    urlcolor=blue,         
    filecolor=black
}

\usepackage[nameinlink,capitalise]{cleveref}
\crefname{section}{Sect.}{Sects.}
\Crefname{section}{Section}{Sections}
\crefname{figure}{Fig.}{Figs.}
\Crefname{figure}{Figure}{Figures}
\crefname{equation}{Eq.}{Eqs.}
\Crefname{equation}{Equation}{Equations}
\crefname{table}{Table}{Tables}
\crefname{appchapter}{Appendix}{Appendices}
\Crefname{appchapter}{Appendix}{Appendices}

\begin{document}


   \title{The eROSITA X-ray luminosity function of \\ active galactic nuclei}

%

   \author{W. Roster\inst{1}\fnmsep\thanks{Corresponding author: wroster@mpe.mpg.de}
        \and J. Buchner\inst{1}
        \and M. Salvato\inst{1,2}
        \and R. Shirley\inst{1}
        \and A. Merloni\inst{1}
        \and T. Dwelly\inst{1}
        \and J. Aird\inst{3}
        \and A. Georgakakis\inst{4}
        \and P. Boorman\inst{1}
        \and Z. Igo\inst{5}
        \and M. Brusa\inst{6,7}
        \and W. N. Brandt\inst{8,9,10}
        \and B. Trakhtenbrot\inst{1,2,11}
        \and B. Laloux\inst{1,12}
        \and P. Baldini\inst{1}
        \and C. Andonie\inst{1}
        \and C. Aydar\inst{1}
        \and C. Ricci\inst{13,14}
        \and S. F. Anderson\inst{15}
        \and P. Chakraborty\inst{16}
        \and R. J. Assef\inst{13}
        \and D. P. Schneider\inst{7}
        \and E. Kyritsis\inst{1}
        \and M. Kluge\inst{1}
        \and E. Bulbul\inst{1}
        \and K. Nandra\inst{1}
        \and J. Weller\inst{1,2,17}}

   \institute{Max-Planck-Institut f\"ur extraterrestrische Physik, Giessenbachstr. 1, 85748 Garching, Germany
   \and Exzellenzcluster ORIGINS, Boltzmannstr. 2, D-85748 Garching, Germany
   \and Institute for Astronomy, University of Edinburgh, Royal Observatory, Blackford Hill, Edinburgh EH9 3HJ, UK
   \and Institute for Astronomy \& Astrophysics, National Observatory of Athens, V. Paulou \& I. Metaxa, 11532, Greece
   \and European Space Agency (ESA), European Space Astronomy Centre (ESAC), Camino Bajo del Castillo s/n, E-28692 Villanueva de la Cañada, Madrid, Spain
   \and Dipartimento di Fisica e Astronomia, Università di Bologna, via Gobetti 93/2, I-40129 Bologna, Italy
   \and INAF—Osservatorio di Astrofisica e Scienza dello Spazio di Bologna, via Gobetti 93/3, I-40129 Bologna, Italy
   \and Institute for Gravitation and the Cosmos, The Pennsylvania State University, University Park, PA, 16802, USA
   \and Department of Astronomy and Astrophysics, The Pennsylvania State University, University Park, PA 16802, USA
   \and Department of Physics, 104 Davey Laboratory, The Pennsylvania State University, University Park, PA 16802, USA
   \and School of Physics and Astronomy, Tel Aviv University, Tel Aviv 69978, Israel
   \and INAF-Osservatorio Astronomico di Capodimonte, Via Moiariello 16, 80131 Napoli, Italy
   \and Department of Astronomy, University of Geneva, ch. d’Ecogia 16,
   1290, Versoix, Switzerland
   \and Instituto de Estudios Astrof\'isicos, Facultad de Ingenier\'ia y Ciencias, Universidad Diego Portales, Santiago, Chile
   \and Department of Astronomy, University of Washington, Box 351580, Seattle, WA 98195, USA
   \and Center for Astrophysics | Harvard \& Smithsonian, 60 Garden St., Cambridge, MA 02138, USA
   \and Universit\"ats-Sternwarte, Ludwig-Maximilians Universit\"at M\"unchen, Scheinerstr. 1, 81679 M\"unchen, Germany}

   \date{Received September 30, 20XX}

     \abstract
{The X-ray luminosity function (XLF) of active galactic nuclei (AGN) provides an observational probe of the growth of supermassive black holes (SMBHs) across cosmic time. With its large survey grasp, Spectrum Roentgen Gamma (SRG)/eROSITA samples the luminosity--redshift plane with a depth--area balance complementary to pencil-beam surveys, providing the volume needed to detect luminous AGN previously limited by small-number statistics.}
{We measure the soft XLF, leveraging an eROSITA sample spanning approximately eight orders of magnitude in luminosity and reaching out to $z\simeq6$. This enables us to study luminosity-dependent evolution with improved constraints, and to infer both the SMBH accretion history and optical/UV missed AGN population.}
{Combining X-ray-selected AGN from two eROSITA-DE survey components, we infer the XLF using a forward-folding Poisson point-process likelihood. We introduce a new redshift-dependent smoothly broken power-law parameterisation in which all XLF model parameters are allowed to evolve continuously with redshift.}
{While broadly consistent over common luminosity and redshift ranges, we find lower space densities for moderately and very luminous AGN at low redshift, while the abundance is higher than previously found at the highest redshifts. Comparisons to optical/UV quasar LFs converted to rest-frame $2\!-\!10\,\mathrm{keV}$ show that the UV-missed fraction decreases with luminosity. In the most luminous bin, the missed fraction increases with redshift, suggesting a strong increase of the obscured fraction towards $z\sim6$. Integrating the XLF yields a black-hole accretion-rate density peaking at $z\simeq1.5$, with the corresponding cumulative black-hole mass density indicating $\sim80^{+11}_{-23}\%$ obscured growth relative to locally-inferred BH mass estimates derived from scaling relations and missed by the soft X-ray selection of our sample.}
{With this work, we release the eROSITA DR2 AGN catalogue, including counterparts and their redshift information, providing a resource for follow-up studies and survey planning. We then discuss how these results can inform future spectroscopic, photometric, and X-ray survey strategies aimed at improving AGN demographic constraints.}


   \maketitle

\nolinenumbers

\section{Introduction}

The luminosity function (LF) of active galactic nuclei (AGN) quantifies the comoving number density of accreting supermassive black holes (SMBHs) as a function of luminosity and redshift \citep[e.g.,][]{Page_1997, schmidt_1999, Miyaji2000, Ueda2003, Hasinger_2005, Treister_2009}. It is therefore a key observable for tracing SMBH growth and the co-evolution with their host galaxies across cosmic time \citep[e.g.,][]{Kormendy1995, Cowie_1996, Magorrian1998, Ferrarese_2000, Gebhardt2000, Marconi_2004, Merloni_2008, Gültekin_2009, Fontanot_2009, Kormendy2013}. Previous studies of the AGN population have primarily relied on samples selected in the optical \citep[e.g.,][]{Schmidt_1968, Schmidt_1983, Boyle2000, Croom2004}, ultraviolet \citep[UV, e.g.,][]{Fan_1999, Masters_2012, Mcgreer_2013, Manti_2017, Kulkarni2019}, infrared \citep[IR, e.g.,][]{Assef_2011, Lacy_2015}, and X-ray bands \citep[e.g.,][]{Maccacaro_1991, Lafranca2005, Ebrero2009, Aird2010, Buchner_2015, Ananna_2022, Pouliasis_2025, BH_2025}, each tracing different components of the AGN population. Across these regimes, X-ray selection provides a particularly clean route to identifying accretion, being less affected by host-galaxy dilution than optical/UV methods and less contaminated by star formation than IR selections, although it remains incomplete for the most heavily obscured sources \citep{Cardamone_2008, Burlon_2011, Delmoro_2016, Hickox_2018, Mountrichas_2020, Brandt_2022}.

Early studies of the AGN space density, commonly parametrise the LF as a double power law combined with either pure luminosity evolution \citep[PLE; e.g.,][]{Maccacaro_1983, Maccacaro_1984, Page_1996, Jones_1997, Boyle2000, Barger2005}, in which the LF shifts in luminosity while retaining its shape, or pure density evolution \citep[PDE; e.g.,][]{Schmidt_1968, Wolf_2003}, in which only the normalisation evolves with redshift. Such prescriptions were soon found to be too restrictive to reproduce the observed AGN population \citep[e.g.,][]{Hasinger_2005, Aird2010, Volonteri_2012, Kalfountzou_2014, Georgakakis2015, Miyaji_2015}, motivating more flexible forms such as luminosity and density evolution \citep[LADE;][]{Yencho_2009, Aird2010} and luminosity-dependent density evolution \citep[LDDE;][]{Schmidt_1983, Miyaji2000, Hasinger2005, Lafranca2005, Silverman2008, Ueda_2014, Peca_2023}. More recent work has further relaxed the assumed LF shape, allowing highly flexible descriptions  \citep[e.g.,][]{Aird_2015, Buchner_2015, Ananna_2019, Laloux_2023}. For instance, \citet{Aird_2015} modelled the X-ray LF (XLF) as a smoothly bending power law whose parameters evolve polynomially with redshift.

Despite these advances, XLF model predictions still diverge substantially in poorly sampled regions of the luminosity--redshift plane, particularly at the bright and high-redshift ends where measurements remain limited by small-number statistics \citep[e.g.,][]{Hasinger2005, Aird2010, Ueda_2014, Aird_2015, Fotopoulou_2016}. Previous XLF measurements, therefore, relied on combining complementary survey tiers \citep{Ueda2003, Ueda_2014, Aird2010, Vito_2014, Buchner_2015, Miyaji_2015}. Shallow, wide-area surveys provide the large sky coverage necessary to sample the cosmological volumes over which the rarest luminous high-redshift AGN can be found, reaching populations with space densities of order $\sim 1\, \rm{Gpc^{-3}}$, albeit at the expense of limited sensitivity and angular resolution \citep[e.g,][]{Wolf2023}. Conversely, deeper pencil-beam observations probe fainter, Seyfert-like sources over small fields but to high redshift, yielding complementary constraints on the numerous low-luminosity population. Although their good positional accuracy facilitates reliable cross-identification with other photometric bands, many faint X-ray AGN are also intrinsically faint in the optical, additionally hampering follow-up with typically patchy spectroscopic coverage. Likewise, the varying availability of auxiliary imaging data can affect the quality of multiwavelength counterparts and their photometric redshift (photo-$z$) estimates \citep[e.g.,][]{Barger2005, Salvato_2009, Cardamone_2010, norris2019, Barro_2019, Luo_2017}. However, such federated data sets require careful homogenisation of selection functions, instrument spatial resolutions and energy responses, backgrounds, imperfections in instrument cross-calibrations, as well as X-ray spectral fitting approaches and assumptions \citep{Tsujimoto_2011}. Consequently, no single survey has simultaneously provided the depth, area, and redshift completeness needed to sample all relevant regions of the XLF with satisfactory statistical power \citep{Peca_2023, BH_2023, BH_2025}.

\begin{figure}[t!]
    \centering
    \includegraphics[width=1\linewidth]{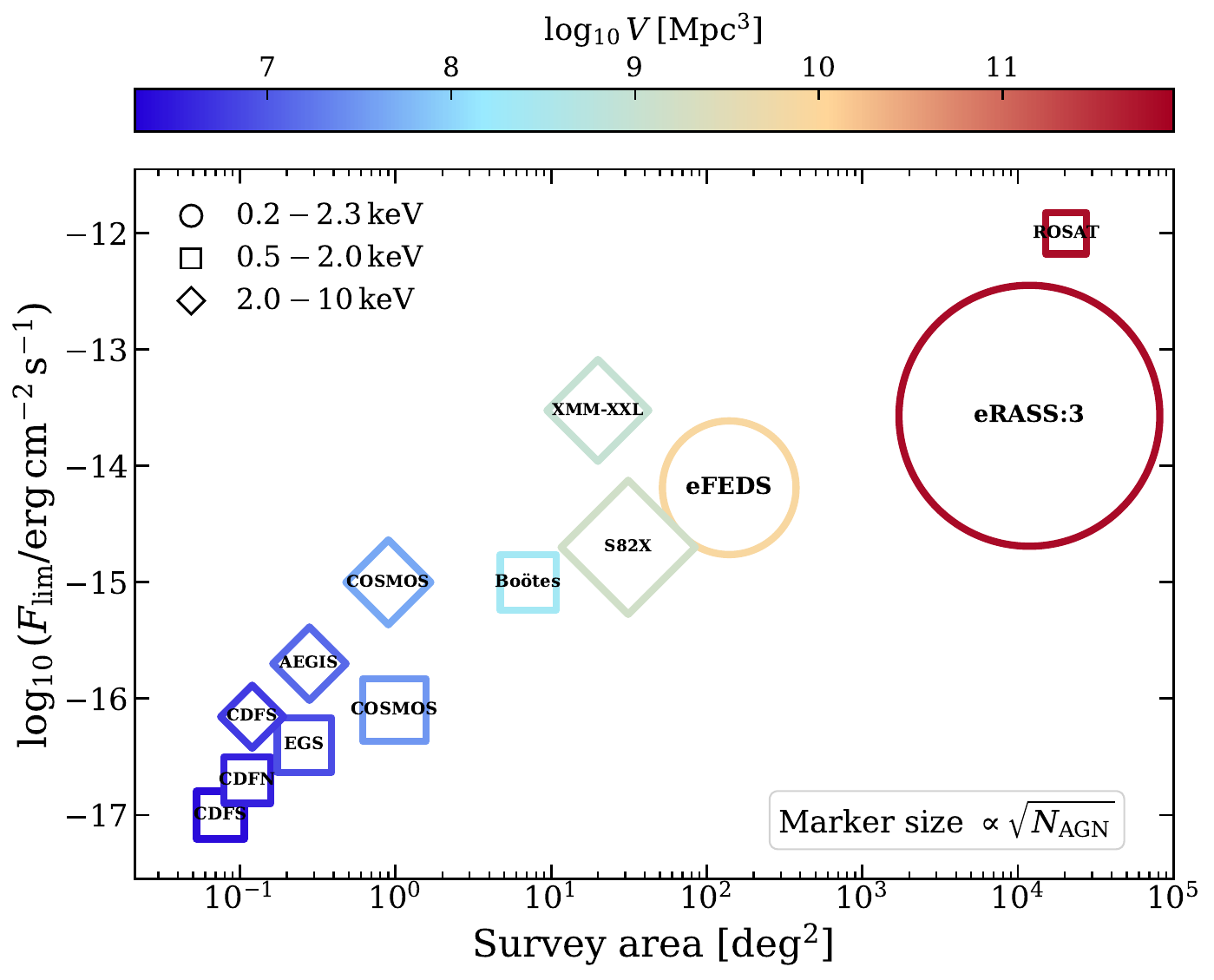}
    \caption{Illustrative comparison of demographic leverage for selected X-ray surveys. Marker size scales with the square root of the number of AGN in each X-ray survey used in literature XLF studies, while the colour indicates the geometric comoving volume, $V$.}
    \label{fig:area_depth}
\end{figure}

Against this backdrop, the extended ROentgen Survey with an Imaging Telescope Array \citep[eROSITA,][]{Predehl2021} aboard Spectrum-Roentgen-Gamma \citep{Sunyaev_2021}, bridges the gap between all-sky surveys such as \textit{ROSAT} \citep{Voges1999, Boller_2016}, MAXI \citep{Hiroi_2011}, and \textit{Swift}-BAT \citep{Oh_2018} and pencil-beam surveys conducted with \textit{Chandra} \citep[e.g.,][]{Evans_2024} and \textit{XMM-Newton} \citep[e.g.,][]{Webb2020}, by providing a new depth--area regime for X-ray AGN studies \citep[see \cref{fig:area_depth},][]{merloni2012erosita, merloni2024}. Compared with previous all-sky surveys, eROSITA combines improved depth with reduced positional uncertainty to enhance counterpart identification and redshift completeness \citep{Kolodzig_2013, Salvato25, Ramos_2026}. As such, we are now in a position to place significantly tighter constraints on the functional form of AGN evolution, to, for example, improve cosmological simulations and semi-analytic models that track SMBH seeding, accretion, and feedback \citep[e.g.,][]{Di_Matteo_2008, Degraf_2010, Fanidakis_2012, Somerville_2015, Habouzit_2022, Weinberger_2025}, or contrain population models like such following the Soltan argument \citep[e.g.,][]{Hopkins_2007, Aird_2015, Inayoshi_2024}.

With an average point-source sensitivity of 
\mbox{$F_{\rm X\,(0.2-2.3\,keV)} \gtrsim 2.7\times10^{-14}\,\mathrm{erg\,cm^{-2}\,s^{-1}}$},
DR2 provides a high-purity soft X-ray view of the AGN population, while primarily detecting AGN with low to moderate obscuration. It is therefore biased against systems for which photons in this energy range are absorbed or scattered along the line of sight \citep{Liu_2022}. The impact of obscuration depends on luminosity, redshift, and spectral shape, but Compton-thin (CTN) ($N_{\rm H} \sim 10^{22}-10^{24}\,\mathrm{cm^{-2}}$) absorption can already suppress the observed soft-band flux substantially. Observational studies indicate that the fraction of CTN AGN ($N_{\rm H} \sim 10^{22}-10^{24}\,\mathrm{cm^{-2}}$) relative to all detected AGN, declines with increasing X-ray luminosity \citep{Ueda2003, Steffen2003, DC2008, Hasinger_2008, Ebrero2009, Burlon_2011, Ueda_2014, Buchner_2015}, implying that incompleteness due to intermediate obscuration is likely less pronounced in the luminosity regime probed by eROSITA. By contrast, above $N_{\rm H} \gtrsim 10^{24}\,\mathrm{cm^{-2}}$, the large column densities imply substantial optical depths to Compton scattering, so photons can undergo multiple interactions while traversing the absorbing material, progressively reducing their energies to levels at which photoelectric absorption becomes increasingly important. Consequently, a large fraction of Compton-thick (CTK) AGN remain undetected beyond the local Universe, even at high luminosities \citep{Akylas_2012, Ricci_2015, Buchner_2015, Vito_2018, Lanzuisi_2018, Ananna_2019}. This missing population is likely significant, since absorption-corrected XLFs and cosmic X-ray background (CXB) synthesis models require substantial CTK contributions, in some cases approaching or exceeding $\sim50\%$ of the intrinsic AGN population \citep[e.g.,][]{Gilli_2007, Fiore_2009, Buchner_2015, Ananna_2019, Torres_alba_2021, Laloux_2023, Civano_2024}. As such, a substantial fraction of SMBH growth likely remains hidden in sources that are strongly suppressed in the observed eROSITA soft X-ray band. At $z\gtrsim3$, however, the observed $0.2{-}2.3\,\mathrm{keV}$ band corresponds to rest-frame $\sim1{-}10\,\mathrm{keV}$ energies, where absorption by CTN columns is less severe. We use this shift in rest-frame bandpass, together with the large eROSITA source statistics, to constrain the high-redshift XLF and compare it to UV-selected LFs. These predominantly trace unobscured, accretion-disc–dominated AGN, enabling the assessment of the fraction of X-ray-selected AGN missed by UV selection \citep{Wolf_2003, CW_2013, Vito_2014, Georgakakis2015, Georgakakis_2017, Kulkarni2019, Lusso_2023}. 

This paper is structured as follows: \Cref{sec2} presents the eROSITA X-ray sample, including the selection function and redshift determination. In \cref{sec3} we present the XLF function and fitting procedure, while \cref{sec4} gives an overview of the XLF's evolutionary trends. \Cref{sec5} discusses the findings followed by \cref{sec6}, which concludes with a summary. We express magnitudes in the AB system \citep{oke_1983} and assume cosmological parameters of $H_0=70\,$km\,s$^{-1}$\,Mpc$^{-1}$, $\Omega_{\rm M}=0.3$, and $\Omega_{\Lambda}=0.7$.


\section{Data}
\label{sec2}

eROSITA DR2 contains roughly $2$ million X-ray detections in the $0.2$--$2.3\,\mathrm{keV}$ band, almost tripling the extragalactic surface density with respect to eROSITA DR1 \citep{merloni2024}, forming the basis of the extragalactic source catalogue released with this paper. In this section, we first summarise the multiwavelength counterpart identification and classification, presented in \cite{Ramos_2026}. Then, we describe the photometric classification and redshift association for the entire DR2 sample, before defining the subset of sources used for the XLF analysis in this work.

\subsection{X-ray-selected parent sample}

Starting from DR2, we apply a set of survey- and source-level quality cuts to define a high-purity sample of reliable X-ray detections (refer to \cref{tab:downselection} for the successive downselection steps). At the catalogue level, we first restrict the sample to point-like detections by requiring
$
\texttt{EXT\_LIKE} = 0,
$
thereby excluding extended sources, which are predominantly associated with galaxy clusters or diffuse emission rather than the point-like AGN population targeted in this work. We further remove sources flagged during detection and processing. These quality flags identify detections that may be affected by instrumental artefacts, edge effects, problematic background modelling, or other pipeline-related issues that could compromise the reliability of the X-ray source parameters \citep[see Table 2 in][]{Ramos_2026}.

\subsection{Multiwavelength counterparts}
\label{sec:multi_ctp}

For extragalactic science, the DR2 catalogue hinges on redshift completeness, ideally from spectroscopy. Since these are obtained from targeted optical observations rather than from the X-ray detections themselves, sources must first be associated with their multiwavelength counterparts (CTPs). This step is non-trivial for wide-area X-ray surveys, whose positional uncertainties make simple nearest-neighbour matching prone to misidentifications, especially in crowded regions \citep[e.g.,][]{Roster_2025_euclid}. Following \cite{Salvato_2022, Salvato25}, \cite{Ramos_2026} therefore identify optical CTPs in the 10th data release of the DESI Legacy Imaging Surveys \citep[LS10,][]{Dey2019} using the Bayesian cross-matching algorithm \texttt{NWAY} \citep[][]{Salvato2018, Buchner_2021}. In addition, they also provide a Galactic/extragalactic classification based on \cite{Salvato25}. We select the most probable association for each X-ray source by requiring
$
\texttt{NWAY\_ncat} > 1
\quad \mathrm{and} \quad
\texttt{NWAY\_match\_flag}==1\,.
$
Finally, to retain extragalactic sources only, we select
$
\texttt{class\_gal\_exgal} > 0\,.
$
Sources failing this criterion are not necessarily independently confirmed Galactic objects, and a small residual fraction of genuinely extragalactic sources may consequently be excluded, with an estimated false-negative rate of $\lesssim0.15\%$ (Merloni et al., in prep.).

\subsection{Redshift information}

When available, spectroscopic redshifts (spec-$z$s) are assigned by positionally cross-matching the optical CTPs of eRASS:3 point sources to a homogenised compilation of public spectroscopic measurements within $1\,\mathrm{arcsec}$. This compilation builds on existing large spectroscopic collections \citep[e.g.,][]{Kluge_2024,Igo_2026}, but was substantially revised and enriched for the present work. We retained only highly reliable redshifts, incorporated additional AGN-focused spectroscopic catalogues, and applied survey-specific quality cuts to homogenise the reliability of the retained redshifts. The final compilation contains $\sim20$ million spec-$z$s drawn from major wide-area surveys, including multiple generations of the Sloan Digital Sky Survey \citep[SDSS;][]{Bowen_1973, Gunn_2006, Smee_2013}, including visually inspected entries up through the latest releases DR19 \citep{sdss_dr19} and DR20 \citep{sdss_dr20}, the Quaia catalogue \citep{Storey-Fisher_2024} for sources with $\texttt{PQSO}$ or $\texttt{PGAL} \gtrsim 0.8$, as well as other spectroscopic surveys such as the 2dF QSO catalogue \citep{Croom_2009}, the 6dF Galaxy Survey \citep{Jones_2009}, and the first data release of DESI \citep{Desi_dr1}, among others not listed here. For sources without spectroscopy, we rely on photo-$z$s obtained with the machine-learning algorithm \texttt{CIRCLEZ} \citep{Saxena24}, shown to provide more reliable point estimates and PDFs than comparable template fitting approaches for X-ray-selected AGN through the use of multi-band aperture colour information from LS10.

For the sources with spec-$z$ coverage, the photo-$z$ performance ($\sigma \simeq 0.08, \eta \simeq 0.17$, see \cref{fig:redshifts}) matches or exceeds that of previous X-ray studies while being a substantially larger sample \citep[e.g.,][]{Aird_2015, Pouliasis_2025}, as summarised in \cref{tab:downselection}. With this paper, we release all the above-mentioned redshift information, including $\sim 400$k sources with spec-$z$ as well as photo-$z$ estimates for $\sim 1.2$ million DR2 sources (see \cref{tab:downselection}).

\begin{figure}[t!]
    \centering
    \includegraphics[width=1\linewidth]{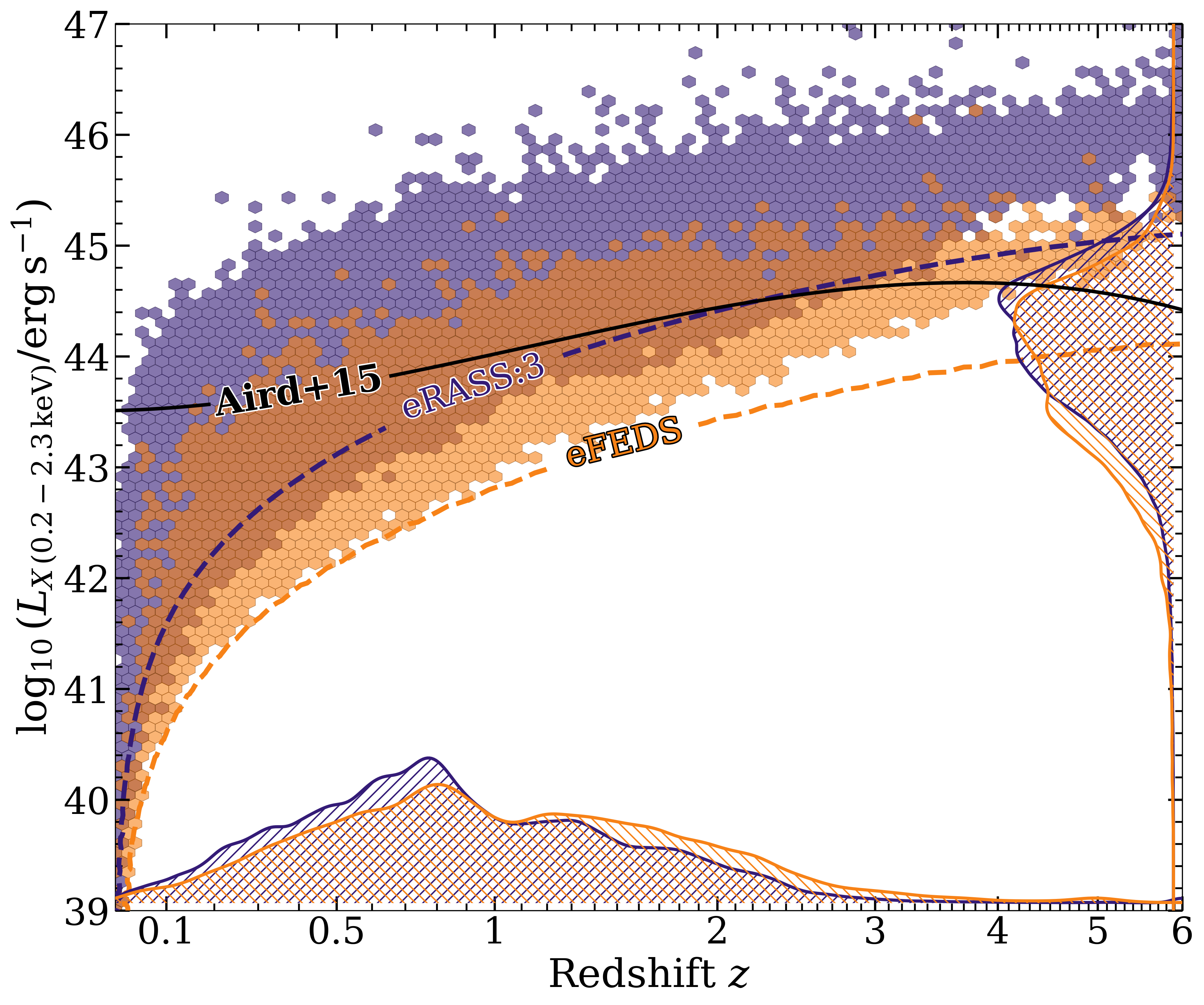}
    \caption{Rest-frame $0.2 - 2.3$ keV luminosity--redshift distribution of the sources used in the XLF analysis. Purple and orange hexagons show the occupied regions of the DR2 and eFEDS samples, respectively. Dashed lines indicate the $1\%$ effective-area limit for the two samples. The solid black curve indicates the break-luminosity evolution $L_\star(z)$ according to the \cite{Aird_2015} soft-band selected flexible double power-law (FDPL) model. The marginal density curves show the corresponding redshift and luminosity distributions.}
    \label{fig:Lxz}
\end{figure}

\subsection{AGN XLF sample}
\label{sec:xlf_sample}

To measure the XLF over the widest $L_{X}-z$ range possible, we adopt a ``wedding-cake'' strategy that combines two complementary eROSITA datasets. While the wide-area DR2 sample provides the cosmological volume needed to constrain rare and luminous AGN, it has limited leverage on lower-luminosity sources once they approach the DR2 sensitivity boundary. We therefore complement DR2 with the eROSITA Final Equatorial-Depth Survey \citep[eFEDS;][]{Brunner_2022, Salvato_2022, Aydar_2025}, a contiguous performance-verification field, approximately four times deeper than the DR2 and slightly deeper than the depth expected for eRASS:8. We show the rest-frame $0.2-2.3$ keV luminosities plotted as a function of redshifts for both the DR2 and eFEDS samples in \cref{fig:Lxz}. 

\subsubsection{Downselection: eFEDS}
\label{sec:efeds_downselection}

We start from the point-source catalogue of 27\,369 multiwavelength counterparts and retain only extragalactic objects by requiring
$
\texttt{CLASS\_CTP} \geq 0\,,
$
and
$
\texttt{CTP\_quality} > 1\,,
$
consistent with \cref{sec:multi_ctp} and reducing the sample to  22\,079 sources (see \cref{tab:downselection}). Combined with its excellent spectroscopic coverage of above 70\%, eFEDS therefore provides a key ingredient in probing the XLF around and below the knee, where the survey depth permits. In the likelihood analysis of \cref{sec:likelihood}, eFEDS sources are treated as a separate survey layer rather than being merged geometrically into the DR2 footprint. Both layers are therefore treated as independent, with its own effective area and selection function, while assuming a single underlying XLF for both.

\subsubsection{Downselection: DR2 survey footprint}
\label{sec:footprint}

Starting from the DR2 catalogue described above, we restrict the data used in the XLF analysis to a DR2 footprint with reliable ancillary coverage in addition to excluding eROSITA tiles with substantial foreground contamination or Galactic absorption \citep{merloni2024}. Specifically, tiles are required to satisfy all of the following criteria:
\begin{itemize}
    \item Gaia DR2 stellar density\footnote{\citet{Gaia_2018}} $< 3\times10^{4}\,\mathrm{deg}^{-2}$,
    \item Foreground reddening\footnote{\citet{Planck_2013}} $\mathrm{E}(\mathrm{B-V}) < 0.3$,
    \item Galactic column density\footnote{\citet{HI4PI_2016}} $N_{\rm H} < 10^{21}\,\mathrm{cm}^{-2}$.
\end{itemize}
This footprint selection retains 1422 of the 1763 DR2 tiles, each covering $3.6 \times3.6\,{\rm deg^{2}}$ (see \cref{tab:downselection}). Since the eFEDS field is included as a separate deep survey layer, the 28 tiles overlapping the eFEDS field are also excluded from the DR2 component to prevent including the same sources twice, as they are not independent measurements (see \cref{tab:downselection}). 

\begin{table*}[!t]

\centering
\small
\caption{Summary of the DR2 catalogue construction and XLF sample selection.}
\label{tab:downselection}

\begin{tabular}{llrr|ccccc}
\toprule
\textbf{Sample} &
\textbf{Selection step} &
\textbf{Removed} &
\textbf{Remaining} &
\multicolumn{5}{c}{\textbf{Sources by redshift bin}} \\
\cmidrule(lr){5-9}
& & & &
$0<z\leq1$ &
$1<z\leq2$ &
$2<z\leq3$ &
$3<z\leq4.5$ &
$z>4.5$\\
\midrule

\multicolumn{8}{l}{\textit{eROSITA all-sky survey DR2}} \\
\midrule
& X-ray detections
& --        & 1\,975\,540 & -- & -- & -- & -- & --\\

& X-ray point-like
& 63\,796 & 1\,911\,744 & -- & -- & -- & -- & --\\

& No flags
& 28\,009  & 1\,883\,735 & -- & -- & -- & -- & --\\

& Within LS10 area\textsuperscript{a}
& 406\,703  & 1\,477\,032 & -- & -- & -- & -- & --\\

& Extragalactic
& 178\,471  & 1\,298\,561 & -- & -- & -- & -- & --\\

& W/ photo-$z$\textsuperscript{b}
& --        & 1\,298\,561
& 787\,907 & 352\,868 & 113\,183 & 20\,770 & 23\,833 \\

& W/ spec-$z$
& 886\,943        & 411\,618
& 187\,624 & 175\,923 & 44\,291 & 3\,742 & 38 \\

\midrule
\multicolumn{8}{l}{\textit{XLF component: eROSITA all-sky survey DR2}} \\
\midrule

& Tile selection\textsuperscript{c} 
& 273\,429  & 1\,023\,725 & -- & -- & -- & -- & --\\

& Removal of eFEDS tiles
& 19\,161   & 1\,004\,584 & -- & -- & -- & -- & --\\

& Detection-significance
& 647\,160  & 357\,424    & -- & -- & -- & -- & --\\

& Non-jetted
& 3\,535    & 353\,889    & -- & -- & -- & -- & --\\

& W/ photo-$z$
& --        & 353\,889 
& 222\,127 & 105\,249 & 21\,411 & 3\,494 & 1\,608\\

& W/ spec-$z$
& 168\,532        & 185\,357 
& 106\,179 & 63\,391 & 11\,014 & 768 & 5 \\

\midrule
\multicolumn{8}{l}{\textit{XLF component: eROSITA eFEDS}} \\
\midrule

& \citet{Salvato_2022}
& --        & 27\,369     & -- & -- & -- & -- & --\\

& Extragalactic
& 5\,290    & 22\,079     & -- & -- & -- & -- & --\\

& W/ photo-$z$
& --        & 22\,079
& 11\,714 & 7\,767 & 2\,228 & 385 & 203\\

& W/ spec-$z$
& 6\,030       & 16\,049
& 6\,968 & 6\,993 & 2\,055 & 210 & 5\\

\bottomrule
\end{tabular}

\tablefoot{
\tablefoottext{a}{Sources with multiple
counterparts are repeated;}
\tablefoottext{b}{Point estimates are defined as the dominant mode of the PDF;}
\tablefoottext{c}{see \cref{sec:footprint}.}}

\vspace{0.5em}
\end{table*}

\subsubsection{Downselection: DR2 detection significance}
\label{detection_sig}

To further suppress spurious or marginal X-ray detections, we apply a detection-significance cut based on the aperture-photometry Poisson probability, \texttt{APE\_POIS}, computed with \texttt{apetool}\footnote{eROSITA Science Analysis Software System \citep[eSASS,][]{Brunner_2022}}. This probability quantifies, for each source, the chance of obtaining the observed number of counts, $H$, within the aperture from a Poisson fluctuation of the locally estimated background, $C_{\mathrm{B}}$, via 

\begin{equation}
    {\mathrm{\texttt{APE\_POIS}}} = P(H \geq H_{\mathrm{min}}  | C_{\mathrm{B}})= \sum_{H = H_{\mathrm{min}}}^{\infty} \frac{C_{\mathrm{B}}^{H}e^{-C_{\mathrm{B}}}}{H!}\,,
\end{equation}
where $H_{\mathrm{min}}$ corresponds to the minimum count required for a detection at the adopted false-positive probability, with smaller values denoting more statistically significant detections of X-ray photons \citep{Georgakakis_2008}. Based on this aperture photometry threshold, analytic area curves for each sky tile were created with \texttt{apetool}. Hence, for consistency, we require
$
0 \leq \texttt{APE\_POIS} \leq 4\times10^{-6},
$
thereby removing the largest fraction of sources in respect to all individual down-selection steps, by retaining only sources for which the measured counts are highly unlikely to arise from background fluctuations (see \cref{tab:downselection}). Correspondingly, simulations predict a very low, few per cent-level spurious-detection fraction \citep{Seppi_2022}.

Lastly, we deselect jetted systems, including predominantly blazars and flat-spectrum radio quasars pointing in our line of sight, whose X-ray emission is not primarily tracing radiatively efficient accretion \citep[e.g.,][]{Ricci_2017}, by requiring
$
\texttt{class\_jetted}=0
$
\citep{Salvato25, Hammerich_2026}. We choose not to invoke a strict luminosity cut (e.g., $L_{X} \geq 10^{42}\, \rm{erg\,s^{-1}}$) as even at the lowest luminosities of our sample, X-ray selected sources are dominated by AGN \citep{Aird_2015}.

\begin{figure}[htbp]
    \centering
    \includegraphics[width=1\linewidth]{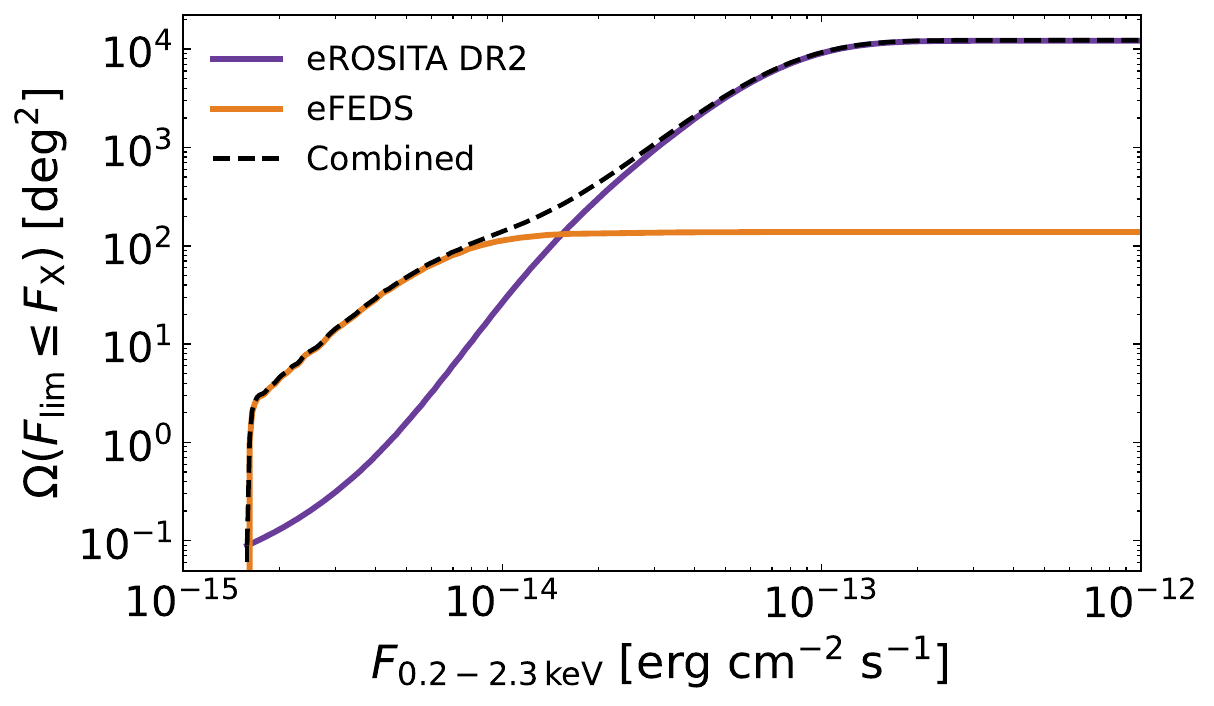}
    \caption{Effective solid angle, $\Omega$, as a function of limiting observed $0.2-2.3\, {\rm keV}$ flux for DR2 and eFEDS \citep{Liu_2022}.  For DR2, the tile-level count-rate curves are summed over the retained XLF footprint. The dashed curve shows their combined coverage.}
    \label{app:selection_function}
\end{figure}

\begin{figure}[t!]
    \centering
    \includegraphics[width=1\linewidth]{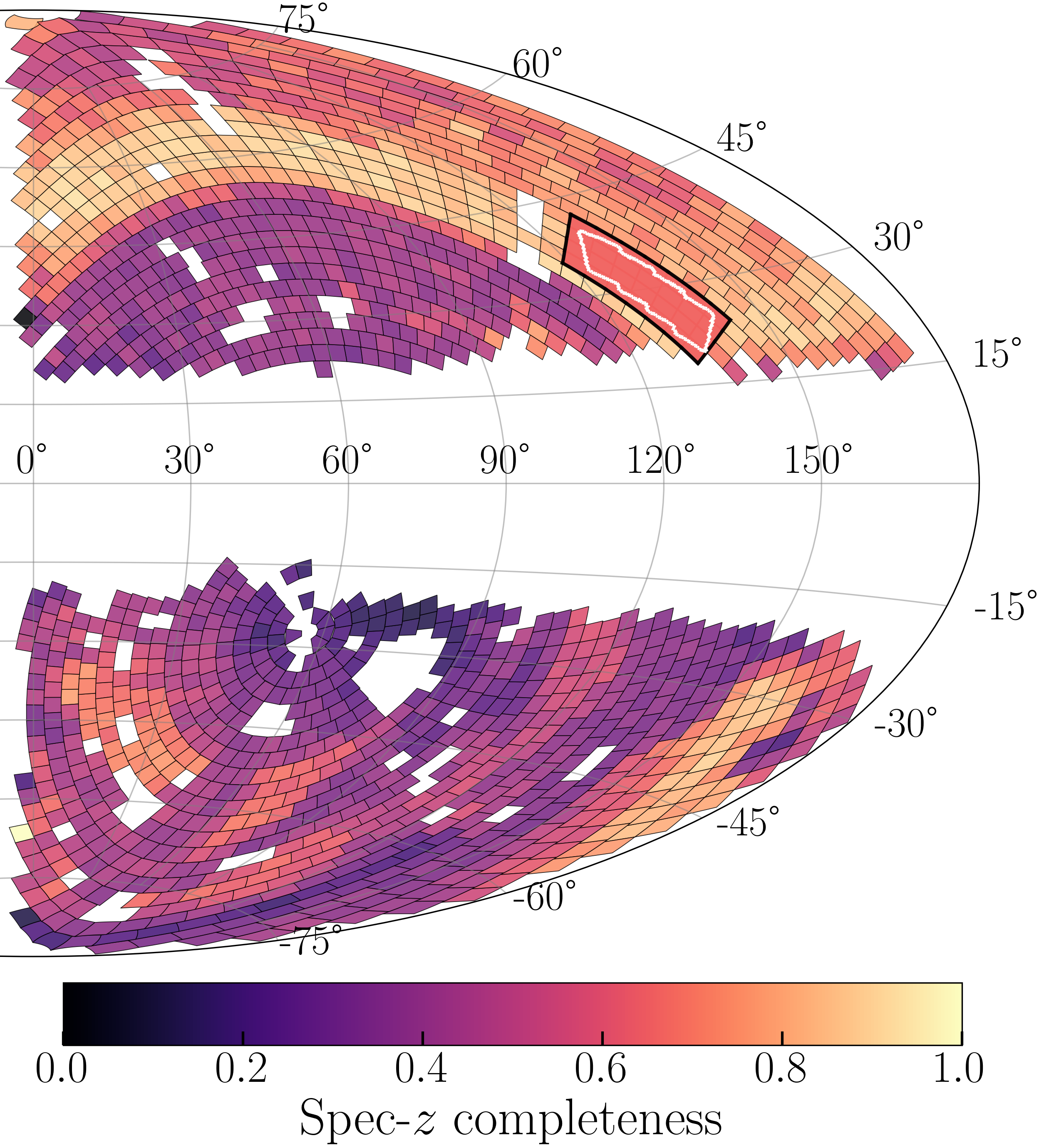}
    \caption{eROSITA-DE western Galactic hemisphere footprint, visualised by the DR2 tiles that enter the XLF sample. Each tile is colored according to its spectroscopic completeness, while uncoloured tiles were excluded by the selection. The tiles covering the eFEDS footprint (outlined in white) are coloured according to their spec-$z$ completeness.}
    \label{fig:tiles}
\end{figure}

The final XLF sample contains $\sim350\,\mathrm{k}$ wide-area DR2 sources, complemented by $\sim22\,\mathrm{k}$ sources from eFEDS, making it orders of magnitude larger than previous AGN samples used for comparable studies (see \cref{tab:downselection} and \cref{fig:area_depth}). \Cref{app:selection_function} shows the respective detection sensitivity of the two eROSITA survey components, while \cref{fig:tiles} shows the retained DR2 tiles, colour-coded by the fraction of sources with spec-$z$. The remaining objects are assigned photo-$z$ estimates, ensuring redshift completeness for virtually the full XLF sample.

\section{Methodology}
\label{sec3}

In this work, we adopt a forward-folding approach to determine the differential co-moving space density of AGN. Under the assumption that the observed data in a survey can be described as Poisson realisations of an underlying population, the likelihood of detecting eROSITA sources as a function of X-ray luminosity and redshift can be formulated as a Poisson point process \citep{Kelly_2008}.

\subsection{Statistical framework and likelihood}
\label{sec:likelihood}

Following works such as \citet{Loredo_2004}, \citet{Aird2010}, \citet{Georgakakis2015}, or \citet{Buchner_2015}, the likelihood separates into a term describing the probability of individually observed sources and a normalisation term that accounts for the expected number of sources given the model. Assuming statistical independence between tiles and surveys, the total likelihood is obtained by summing the Poisson log-likelihood contributions as

\begin{equation}
\ln \mathcal{L}(D \mid \theta)
=
\sum_{k}
\left[
\sum_{i \in k}^{N_{k}}
\ln p(D_i \mid \theta)
-
\lambda_k(\theta)
\right]\,.
\label{eq:loglikelihood1}
\end{equation}
Here, $k$ indexes the individual survey elements entering the likelihood. For DR2, these correspond to the retained sky tiles, while the eFEDS field is treated as an additional effective tile with its own selection function and effective area. This allows the combined DR2+eFEDS likelihood to be written as a single sum over $k$. $D_i$ denotes the observed data for source $i$ out of $N$, $\theta$ represents the XLF model parameters, with $\lambda_k(\theta)$ being the model-predicted numbers of detected sources in tile $k$, obtained by integrating the XLF over the corresponding survey selection functions, $A$, as

\begin{equation}
\lambda(\theta)
=
\int \! \mathrm{d}z \, \frac{\mathrm{d}V}{\mathrm{d}z}
\int \! \mathrm{d}\log L_X \,
A(L_X, z)\,
\phi(L_X, z \mid \theta)\,,
\end{equation}
where $\mathrm{d}V/\mathrm{d}z$ is the comoving volume per solid angle at redshift $z$, $L_{X}$ the observed X-ray luminosity, and $\phi$ the observed XLF model. The likelihood is marginalised over the redshift probability distribution function $p_i(z)$. For sources with spec-$z$, the likelihood is evaluated assuming a delta function at the measured redshift, via

\begin{equation}
\begin{split}
p(D_i \mid \theta)
=
\int \! \mathrm{d}z \, \frac{\mathrm{d}V}{\mathrm{d}z}\,p_i(z)
\int \! \mathrm{d}\log L_X \, \\
p(N_i \mid L_X, z)
\, \phi(L_X, z \mid \theta)\, .
\end{split}
\end{equation}
For sources without a spec-$z$, we propagate the full photometric-redshift probability distribution (PDF) rather than relying on a single point estimate. In particular, \texttt{CIRCLEZ} PDFs have been shown to provide improved reliability compared to traditional template-fitting-based PDFs used in literature \citep{Saxena24}. The likelihood term $p(N_i \mid L_{X}, z)$ accounts for Poisson counting statistics and redshift uncertainties. The observed number of counts $N_i$ is modelled as a Poisson realisation with mean
\begin{equation}
\mu(L_{X}, z) = \mathrm{CR}(L_X, z)\, t_i + b_i,
\end{equation}
where $\mathrm{CR}(L_X,z)$ is the expected source count rate for a source of luminosity $L_X$ at redshift $z$, $t_i$ is the effective exposure time, and $b_i$ is the expected background contribution.

\subsection{Survey selection function}

The measurement of the XLF requires an accurate description of the survey selection function, which quantifies the effective survey volume $A(L_X,z)$ over which sources of a given luminosity and redshift can be detected. In practice, this involves combining the instrumental sensitivity of the survey with assumptions about the spectral properties of X-ray point sources to determine their observable count rates corresponding to a given rest-frame luminosity under the adopted spectral model. In the case of eROSITA, the survey sensitivity varies across the sky due to changes in exposure time, background level, Galactic absorption, and instrumental response. As a result, the effective survey area depends on the minimum detectable count rate at a given tile position, which is why the detection sensitivity as a function of count rate has already been computed with the Poisson aperture assumption, as introduced in \cref{detection_sig}, for each sky tile in DR2. The effective solid angle as a function of limiting observed $0.2-2.3\, {\rm keV}$ flux for the DR2 and eFEDS components is shown in \cref{app:selection_function}.

We model the conversion between count rate and rest-frame luminosity by generating simulated X-ray spectra from a fixed grid of normalisation and redshift, using the spectral fitting package \texttt{XSPEC} \citep{Arnaud_1996}. These assume a typical AGN spectral model: \textit{tbabs*zpowerlw}, consisting of a power-law continuum $F(E) \propto E^{-\Gamma}$, with photon index $\Gamma = 2$ modified by photoelectric absorption fixed to the Galactic absorbing column density appropriate for the full survey footprint \citep{Brunner_2022, Liu_2022}. We retain this approximation because the median and upper $1\sigma$ Galactic $N_{\rm H}$ values across the retained DR2 tiles (see \cref{sec:footprint}) are within 0.1 dex of the adopted eFEDS value. The spectra are convolved with the eROSITA instrumental response to obtain both the count rate and intrinsic X-ray luminosity in the $0.2 - 2.3\, {\rm keV}$ band, allowing us to compute the effective area in luminosity–redshift space (see \cref{fig:Lxz}).

\subsection{LF model and inference}
\label{sec:SBPL}

The commonly adopted smooth double power law

\begin{equation}
    \Phi(L_{\rm X}) \propto \left[ \left(\frac{L_{\rm X}}{L_{\star}}\right)^{\alpha} + \left(\frac{L_{\rm X}}{L_{\star}}\right)^{\beta} \right]^{-1}\,,
\end{equation}
implicitly ties the curvature around $L_{\star}$ to the faint- and bright-end slopes, $\alpha$ and $\beta$. However, given the statistical precision of the eROSITA sample, we instead adopt a generalised smoothly broken power law (SBPL) parameterisation with an explicit width parameter. This allows the sharpness of the transition around the knee to be varied independently while retaining the asymptotic power-law slopes. Mathematically equivalent to the formalism introduced by \citet{Ryde_1999}, the SBPL is written as

\begin{equation}
\begin{split}
\phi(L_{X},z|\,\theta) &= \frac{\mathrm{d}\phi(L_{X}, z)}{\mathrm{d}\, {\rm log}\,L_{X}} \\
&=
C(z)
\left( \frac{L_{X}}{L_\star(z)} \right)^{-\sigma(z)} \times \\
&
\left[
\cosh\!\left(
\frac{\log_{10}(L_{X} / L_\star(z))}{\delta(z)}
\right)
\right]^{-\zeta(z)\,\delta(z)\,\ln 10},
\label{eq:xlf_sbpl}
\end{split}
\end{equation}
where $C$ is the normalisation corresponding to the value of the XLF at the break luminosity, $L_\star$. The parameters $\sigma$ and $\zeta$ are defined in terms of the faint and bright-end slopes, $\alpha$ and $\beta$, as
\begin{equation}
\sigma(z) = \frac{\alpha(z) + \beta(z)}{2},
\qquad
\zeta(z) = \frac{\beta(z) - \alpha(z)}{2},
\end{equation}
while $\delta$ controls the smoothness or width of the characteristic knee between the two power-law regimes. In the asymptotic limits, this form recovers pure power laws with slopes $\alpha$ for $L \ll L_\star$ and $\beta$ for $L \gg L_\star$. In order to reduce parameter degeneracies without loss of generality, we absorb the (arbitrary) pivot defined by \citet{Ryde_1999} by choosing it to coincide with $L_{\star}$. The smoothly connected double power law is recovered when $\zeta\delta {\rm ln}10 = 1$, up to the adopted normalisation at $L_\star$.

\citet{Aird_2015} described the redshift dependence of their XLF parameters with polynomials, finding broadly bending-power-law-like behaviour for several quantities. However, polynomial forms can become unstable when extrapolated beyond the redshift range strongly constrained by the data. In particular, the evolution of $L_*$ in \citet{Aird_2015} exhibits a reversal at high redshift \citep[e.g.,][]{Caplar_2015, Shen_2020}. By contrast, \citet{Buchner_2015} found evidence that the knee evolution saturates towards high redshift. Based on these considerations, utilising flexible parameterisation but allowing saturation, we adopt an SBPL for the evolution function of each parameter,

\begin{equation}
    p(z) \in 
    \left\{
    L_\star(z),\,
    C(z),\,
    \alpha(z),\,
    \beta(z),\,
    \delta(z)
    \right\}\,,
\label{eq:param_evo}
\end{equation}
\noindent listed in \cref{tab:priors}, as
\begin{equation}
\begin{split}
p(z)
&=
p_0
\left[
\frac{1+z}{1+z_{\rm c}}
\right]^{\sigma_p} \times \\
&
\left[
\cosh\!\left(
\frac{
\log_{10}\!\left[(1+z)/(1+z_{\rm c})\right]
}{
\delta_p
}
\right)
\right]^{\zeta_p\,\delta_p\,\ln 10}\,,
\label{eq:param_sbpl}
\end{split}
\end{equation}
where $z_{\rm c}$ denotes the break redshift. Hence, this approach provides a parameter-specific, continuous, and physically motivated description of the XLF evolution that allows simultaneous changes in $C$, $L_{\star}$, slopes, and knee.

We determine the best-fitting XLF parameters by maximising the Poisson point-process likelihood described above. In practice, we minimise the negative log-likelihood of the SBPL model using \texttt{Nelder-Mead} optimisation. Uniform priors are adopted for all parameters, with bounds chosen to encompass physically plausible values while excluding regions of parameter space that lead to numerical instabilities (see \cref{tab:priors}). The resulting parameter vector, $\hat{\theta}$, defines the fiducial best-fitting XLF model used throughout the analysis. 

Because the SBPL parameters are not independent, we find, similar to \citet{Caplar_2015}, substantial covariance between the normalisation $C$, the break luminosity $L_\star$, the asymptotic slopes, and the knee width $\delta$. Broad variations in one parameter can be partially compensated by changes in the others, leading to similar likelihoods for distinct sets of XLF parameters. In particular, changes in $\delta$ modify the luminosity range over which the faint and bright-end slopes are constrained. Given the model's dimensionality, the adopted priors provide a weakly informative regularisation that stabilises the inference in regions of parameter space poorly constrained by the data. However, given the survey quality, we prefer to use independent prior constraints to re-assess all parameters.

To estimate the uncertainties on the inferred parameter evolution, we use a bootstrap procedure where we repeat the full fitting pipeline 500 times on bootstrap realisations of the survey data. Each realisation is constructed by resampling the tiles used in the distributed likelihood calculation. For each bootstrap sample, the SBPL model is re-optimised until convergence. The resulting distribution of best-fitting parameter vectors is then used to define the uncertainty ranges shown in \cref{tab:priors}.

\begin{table*}
\centering
\caption{
Prior ranges and posterior constraints for the redshift-dependent SBPL XLF parameters from \cref{eq:param_evo,eq:xlf_sbpl}.
}
\label{tab:priors}
\renewcommand{\arraystretch}{1.25}
\setlength{\tabcolsep}{5pt}

\begin{tabularx}{\textwidth}{l X c c}
\toprule
\textbf{Parameter} & \textbf{Description} & \textbf{Prior range} & \textbf{Posterior} \\
\midrule

\rowcolor{gray!12}
\multicolumn{4}{l}{\textbf{Break luminosity evolution, $L_\star(z)$}} \\
$L_{\star,\mathrm{norm}}$ 
& Break luminosity at $z=L_{\star,z_c}$ 
& $[1\times10^{43},\,1\times10^{45}]$ 
& $1.15^{+0.03}_{-0.05} \times10^{44}$ \\

$L_{\star,a}$ 
& Low-$z$ slope of $L_\star(z)$ 
& $[3.0,\,8.0]$ 
& $4.73^{+0.09}_{-0.24}$ \\

$L_{\star,b}$ 
& High-$z$ slope of $L_\star(z)$ 
& $[-2,\,4.0]$ 
& $1.79^{+0.13}_{-0.17}$ \\

$L_{\star,z_c}$ 
& Transition redshift 
& $[0.1,\,2.0]$ 
& $0.68^{+0.13}_{-0.11}$ \\

$L_{\star,\delta}$ 
& Transition width 
& $[0.1,\,0.8]$ 
& $0.25^{+0.12}_{-0.05}$ \\

\midrule

\rowcolor{gray!12}
\multicolumn{4}{l}{\textbf{Normalisation evolution, $C(z)$}} \\
$C_{\mathrm{norm}}$ 
& Normalisation at $z=C_{z_c}$ 
& $[1\times10^{-6},\,1\times10^{-4}]$ 
& $2.11^{+0.20}_{-0.20}\times10^{-5}$ \\

$C_a$ 
& Low-$z$ slope of $C(z)$ 
& $[1.5,\,8.5]$ 
& $6.32^{+0.40}_{-1.00}$ \\

$C_b$ 
& High-$z$ slope of $C(z)$ 
& $[-12.0,\,-2.0]$ 
& $-8.80^{+1.00}_{-0.50}$ \\

$C_{z_c}$ 
& Transition redshift 
& $[0.1,\,2.0]$ 
& $0.99^{+0.06}_{-0.13}$ \\

$C_\delta$ 
& Transition width 
& $[0.1,\,0.8]$ 
& $0.46^{+0.02}_{-0.11}$ \\

\midrule

\rowcolor{gray!12}
\multicolumn{4}{l}{\textbf{Faint-end slope evolution, $\alpha(z)$}} \\
$\alpha_{\mathrm{norm}}$ 
& Faint-end slope at $z=\alpha_{z_c}$ 
& $[0.25,\,0.75]$ 
& $0.60^{+0.02}_{-0.09}$ \\

$\alpha_a$ 
& Low-$z$ slope of $\alpha(z)$ 
& $[-4.5,\,-1.5]$ 
& $-1.06^{+0.26}_{-0.10}$ \\

$\alpha_b$ 
& High-$z$ slope of $\alpha(z)$ 
& $[-0.8,\,-0.2]$ 
& $-0.69^{+0.10}_{-0.03}$ \\

$\alpha_{z_c}$ 
& Transition redshift 
& $[0.1,\,2.0]$ 
& $0.24^{+0.10}_{-0.08}$ \\

$\alpha_\delta$ 
& Transition width 
& $[0.1,\,0.8]$ 
& $0.24^{+0.02}_{-0.05}$ \\

\midrule

\rowcolor{gray!12}
\multicolumn{4}{l}{\textbf{Bright-end slope evolution, $\beta(z)$}} \\
$\beta_{\mathrm{norm}}$ 
& Bright-end slope at $z=\beta_{z_c}$ 
& $[1.5,\,3.5]$ 
& $2.15^{+0.01}_{-0.07}$ \\

$\beta_a$ 
& Low-$z$ slope of $\beta(z)$ 
& $[0.0,\,3.0]$ 
& $1.70^{+0.20}_{-0.05}$ \\

$\beta_b$ 
& High-$z$ slope of $\beta(z)$ 
& $[-0.2,\,0.2]$ 
& $0.01^{+0.22}_{-0.06}$ \\

$\beta_{z_c}$ 
& Transition redshift 
& $[0.1,\,2.0]$ 
& $0.18^{+0.04}_{-0.07}$ \\

$\beta_\delta$ 
& Transition width 
& $[0.1,\,0.5]$ 
& $0.27^{+0.01}_{-0.03}$ \\

\midrule

\rowcolor{gray!12}
\multicolumn{4}{l}{\textbf{Knee-width evolution, $\delta(z)$}} \\
$\delta_{\mathrm{norm}}$ 
& Smoothness at $z=\delta_{z_c}$ 
& $[0.1,\,0.5]$ 
& $0.38^{+0.07}_{-0.01}$ \\

$\delta_a$ 
& Low-$z$ slope of $\delta(z)$ 
& $[-0.2,\,0.2]$ 
& $-0.11^{+0.02}_{-0.00}$ \\

$\delta_b$ 
& High-$z$ slope of $\delta(z)$ 
& $[-0.8,\,0.2]$ 
& $-0.29^{+0.02}_{-0.09}$ \\

$\delta_{z_c}$ 
& Transition redshift 
& $[0.1,\,2.0]$ 
& $0.26^{+0.05}_{-0.01}$ \\

$\delta_\delta$ 
& Transition width 
& $[0.1,\,0.5]$ 
& $0.17^{+0.03}_{-0.01}$ \\

\bottomrule
\end{tabularx}

\tablefoot{
All priors are uniform within the quoted bounds. Posterior values are given as the fiducial best-fitting value with asymmetric $1\sigma$ uncertainties estimated from the lower and upper bootstrap ranges. The parameters $C_{\mathrm{norm}}$ and $L_{\star,\mathrm{norm}}$ are given in units of $\mathrm{Mpc^{-3}\,dex^{-1}}$, and $\mathrm{erg\,s^{-1}}$, respectively. }
\end{table*}

\subsection{Model validation and consistency check}

As an independent non-parametric check, we also show binned estimates computed with the \citet{Page_2000} V$_{\rm max}$ estimator using the same combined DR2 and eFEDS selection functions (see \cref{sec:vmax}). These binned estimates are not used as the primary fit statistic, but provide a useful visual comparison to the forward-modelled XLF. To be consistent with most works from literature, we convert our soft-band luminosities to the $2-10$ keV band, assuming a power-law spectrum with photon index $\Gamma$, by

\begin{equation}
\frac{L_{2-{10\,\mathrm{keV}}}}
     {L_{0.2-{2.3\,\mathrm{keV}}}}
=
\frac{
\int_{2\,\mathrm{keV}}^{{10\,\mathrm{keV}}}
E^{1-\Gamma}\,{\rm d}E
}{
\int_{{0.2\,\mathrm{keV}}}^{{2.3\,\mathrm{keV}}}
E^{1-\Gamma}\,{\rm d}E
}\,.
\end{equation}

\noindent For $\Gamma=2$, the corresponding expression is
\begin{equation}
    L_{2-10\,{\mathrm{keV}}} = L_{0.2-2.3\,{\mathrm{keV}}} \times \frac{\ln \left(10\,\mathrm{keV}/2\,\mathrm{keV} \right)}{\ln \left(2.3\,\mathrm{keV}/0.2\,\mathrm{keV} \right)} \approx 0.66\,.
\end{equation}

A more ambitious, fully non-parametric hierarchical reconstruction of the XLF, with independent amplitudes in many luminosity--redshift cells and a smoothness or regularisation prior, is attractive in principle \citep{Buchner_2015}. In practice, however, it is not computationally feasible for the present analysis. The likelihood evaluation requires repeated integration over luminosity, redshift, and the spatially varying eROSITA selection function, including the tile-dependent sensitivity information used for DR2 and the separate eFEDS selection function. In the current pipeline, these selection-function products are already distributed across 83 parallel HPC jobs, each requiring several GB of intermediate data. A fully non-parametric Stan Hamiltonian Monte Carlo implementation would require evaluating this high-dimensional likelihood, and its gradients, for a large number of luminosity--redshift cells at every sampling step, substantially increasing both memory use and wall time \citep{Neal_2011, Hoffmann_2014, Carpenter_2017}. For these reasons, we retain the forward-modelled parametric XLF as the primary inference framework using the \cite{Page_2000} estimates as an independent non-parametric visual diagnostic.

We further validate our XLF inference framework with a forward-modelling test using mock sources drawn from the soft-band-selected FDPL model reported in Table 7 of \cite{Aird_2015}. The mock population is passed through the same survey selection machinery as introduced in \cref{sec2}. For each mock source, we compute the expected count rate using the same spectral assumptions and instrumental response adopted for the selection function, assign exposure times and background levels from the observed DR2 distributions, and generate observed counts from Poisson statistics. Sources are then selected using the same Poisson-based detection criterion as for the XLF sample used in this work. We fit the resulting mock catalogue with our SBPL model and compare the recovered XLF to the input literature model to verify that the SBPL parameterisation recovers the underlying XLF within the posterior uncertainties (see \cref{sec:Aird_recovery}), thereby demonstrating that the adopted SBPL form is flexible enough to effectively describe a more general evolution than LDDE, at least for the context of this paper.


\section{Results}
\label{sec4}

We fit the combined eROSITA DR2 and eFEDS AGN sample with the SBPL model described in \cref{sec3}. Throughout this work, $L_{\rm X}$ denotes the rest-frame luminosity inferred from the observed eROSITA count rate, assuming a $\Gamma = 2$ power-law spectrum corrected for Galactic absorption only. We do not correct for intrinsic absorption as reliable source-by-source constraints on $N_{\rm H}$ or spectral shape are unavailable for the majority of the sample at this time. Intrinsic absorption can therefore affect the inferred XLF in two ways. First, sufficiently absorbed AGN may fall below the soft-band detection threshold and be absent from the sample. Second, absorbed AGN that remain detected have suppressed count rates and are consequently assigned luminosities below their intrinsic, absorption-corrected values, potentially moving them into lower-luminosity bins of the XLF.

\subsection{How does the XLF shape evolve?}

\begin{figure*}[t!]
    \centering
    \includegraphics[width=1\linewidth]{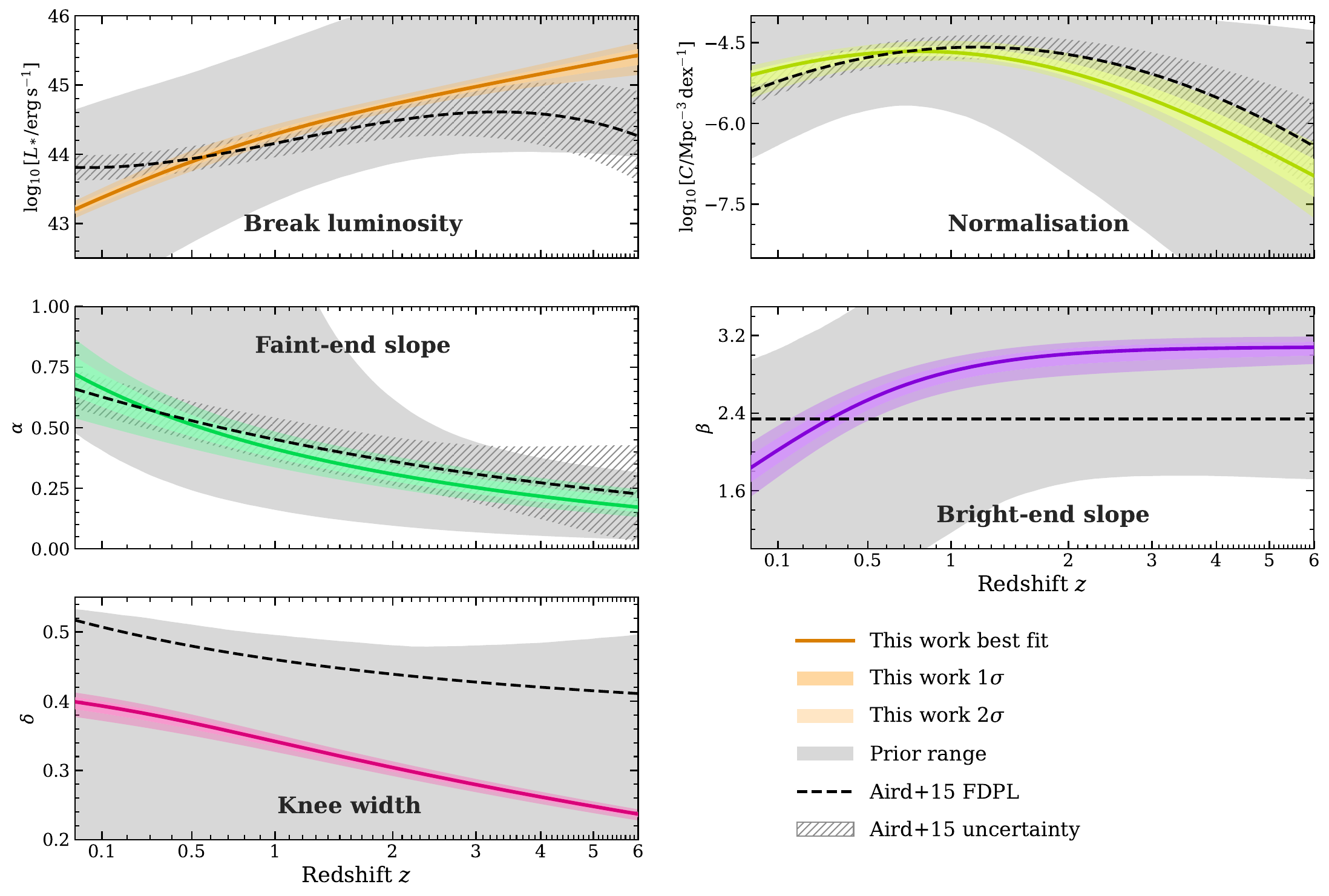}
    \caption{Evolution of the best-fitting SBPL XLF parameters as a function of redshift. The coloured curves indicate the fiducial best-fitting model, with 1$\sigma$ and 2$\sigma$ percentile intervals, relative to the \cite{Aird_2015} soft-band-selected FDPL evolution (shown in black). The grey shaded region indicates the range of parameter evolution permitted by the prior bounds listed in \cref{tab:priors}.}
    \label{fig:param_evo}
\end{figure*}

The redshift evolution of the best-fitting parameters, together with their bootstrap ranges, is shown in \cref{fig:param_evo}. We find significant evolution in all SBPL parameters, with particularly steep redshift dependence at $z\lesssim1$, exceeding the already pronounced low-redshift evolution inferred by previous XLF studies. This includes evolution in the bright-end slope, which is shallower at low redshift ($z\lesssim0.3$) and becomes steeper at higher redshift. In many earlier parameterisations, this slope was held fixed, largely because the bright end was difficult to constrain with limited source statistics at high luminosities and high redshifts \citep[e.g.][]{Hasinger_2005, Ueda_2014, Miyaji_2015, Aird_2015}.

\begin{figure*}[h!]
    \centering
    \includegraphics[width=1\linewidth]{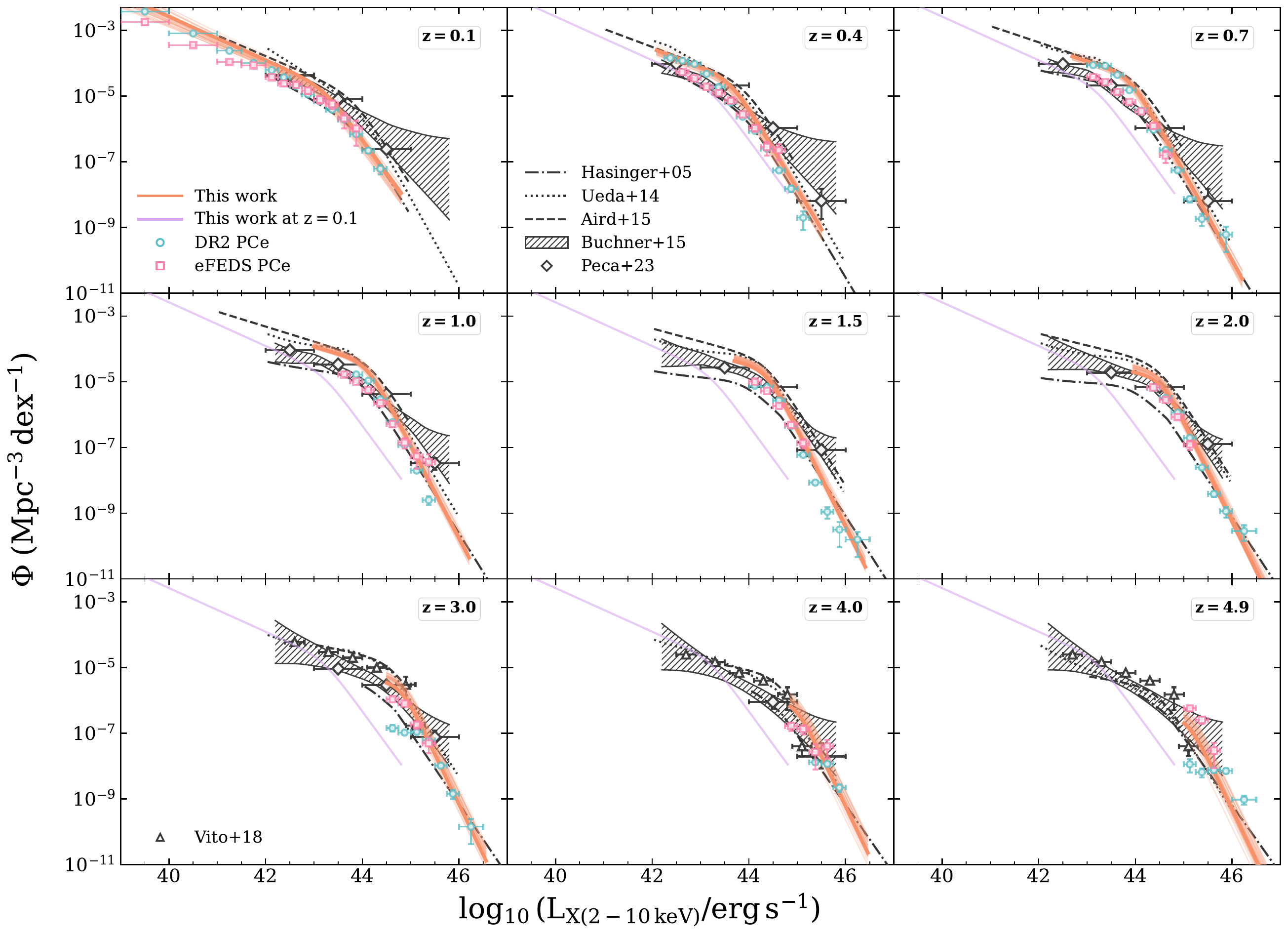}
    \caption{Each panel shows $\Phi$ from \cref{eq:xlf_sbpl} as a function of rest-frame log$_{10}\,(\rm{L}_{2 - 10\,{\rm keV}})$ at the redshift indicated in the upper-right corner. The coloured markers show binned estimates from the DR2 and eFEDS samples computed with the \cite{Page_2000} estimator (PCe) corrected for the selection function. The faint violet curve, repeated in panels other than the lowest-redshift panel, shows this work's XLF fixed at $z=0.1$. External measurements and models (\citealt{Hasinger_2005} LDDE, \citealt{Ueda_2014} LDDE, \citealt{Buchner_2015} credible interval for $N_{\rm H} \leq 10^{22}\,\rm{cm}^{-2}$, \citealt{Aird_2015}  soft-band-selected FDPL, \citealt{Peca_2023} and \citealt{Vito_2018}, both hard-band selected), are shown for comparison.}
    \label{fig:collage}
\end{figure*}

The break luminosity $L_{\star}$ evolves strongly and approximately monotonically over the redshift range probed here (see \cref{fig:param_evo}). Compared to previous parameterisations, our fit favours a lower characteristic break luminosity, 
increasing from log$(L_{\star}) \simeq 43.2$ to log$ (L_{\star}) \simeq 44.1$ by $z \sim 1$, and continues to rise towards log$ (L_{\star}) \simeq 45$, and beyond, through $z \sim 5$, more comparable to trends observed for absorbed sources, at earlier cosmic times \citep{Aird_2015}. However, the flux-limited eROSITA samples miss low-luminosity AGN at high $z$, needed to constrain the evolution in this regime. Moreover, the normalisation reaches its maximum of log$\Phi \simeq -4.7$ at a comparatively lower redshift of $z \sim 0.7$, while declining equally sharply towards higher redshifts, dropping by up to 3 dex by $z\sim5$ (see \cref{fig:param_evo}). These trends should not be interpreted independently: the SBPL parameters are covariant, and changes in $L_{\star}$, $C$, the slopes, and $\delta$ can partially compensate one another while producing similar XLF shapes over the luminosity range constrained by the data. Nevertheless, the combined behaviour indicates that the XLF does not evolve through a pure luminosity or pure density shift alone. Instead, the shape of the LF itself changes with redshift \citep[e.g.,][]{Miyaji2000, Ueda2003, Hasinger_2005, Aird2010}.

Unlike previous XLF studies, in which the transition between the faint and bright-end regimes was set implicitly by the adopted functional form, our parameterisation allows the knee width to vary explicitly with redshift. The best-fitting model favours a relatively sharp transition, with a width of $\sim 0.35$ at $z\simeq 1$ compared to the greater width of $\sim 0.45$ inferred from the \citet{Aird_2015} FDPL form (see \cref{fig:param_evo}). For AGN, the observed knee reflects the convolution of several ingredients, including the galaxy LF, the black-hole mass function, the Eddington-ratio distribution, the luminosity dependence of accretion states, and selection effects \citep[e.g.,][]{Caplar_2015, Tucci_2017,Ananna_2022}. A narrower knee, therefore, suggests that, at a given redshift, the transition between the numerous moderate-luminosity AGN and the rapidly declining luminous-quasar population occurs over a relatively limited luminosity range. One possible physical interpretation of a narrow transition region is that the LF $\delta$ traces a rapid change in the dominant accretion state rather than a gradual redistribution of sources across luminosity. In the radiation-regulated growth scenario of \citet{Ricci_2017, Ricci_2022}, SMBHs spend much of their growth phase embedded in gas and dust, with increasing Eddington ratio accompanied by increasing column density and covering factor. Once the source approaches the effective Eddington limit for dusty gas, radiative feedback can rapidly expel or redistribute the circumnuclear material, producing a short-lived blowout phase followed by a more unobscured state. In this picture, the break in the LF may correspond to the luminosity/Eddington-ratio regime where the AGN population transitions from obscured, fuel-rich growth to feedback-regulated depletion, naturally leading to a relatively sharp $\delta$ if the transition timescale is short. However, as mentioned, at high redshift where the eROSITA sensitivity limit approaches the knee of the XLF, the lowest-luminosity bins can no longer be fully sampled, reducing the leverage on the faint-end shape and potentially biasing the inferred transition width (see \cref{fig:collage}).

A related effect is visible in the evolution of the faint-end slope: while at low redshift the large number of low-luminosity X-ray detections, including an increased fraction of non-AGN emitters, provides strong constraints, the incompleteness of faint sources at higher redshift causes the inferred faint-end slope to deviate from reference trends and become increasingly driven by the adopted parametrisation of the evolution.

In each of the panels of \cref{fig:collage}, the orange curve indicates this work's best-fitting SBPL model in each redshift slice over the luminosity interval directly supported by the data, while the shaded region shows 100 model draws within the corresponding $2 \sigma$ bootstrap uncertainty. We deliberately restrict the displayed main model curve to the data-supported parameter range to avoid visually overemphasising extrapolated parts of the fit. Where the effective survey volume is appreciable, the binned V$_{\rm max}$ estimates are generally well described by the best-fitting model. At the lowest luminosities, the binned points appear to turn over below the smooth XLF. These points lie close to the survey sensitivity boundary, where the effective area decreases rapidly, and the accessible volume becomes small. We therefore assign this trend to selection-limited bins, rather than interpreting it as evidence for a premature downturn in the AGN space density. To illustrate the evolution of the XLF shape, we additionally show the best-fitting low-redshift model at $z=0.1$ in each panel.

\subsection{Comparison to literature}
\label{sec:comp_to_lit}

We also compare our XLF to a set of literature measurements and (non-)parametric models, chosen to be as similar as possible to our selection in terms of energy band and obscuration regime \citep{Hasinger_2005, Ueda_2014, Buchner_2015, Aird_2015, Peca_2023, Vito_2018}. However, this comparison should be interpreted with some caution. For example, the \cite{Buchner_2015} constraints include hard-band-selected sources and are separated by column-density bins, while our baseline fit is based on a soft-band-selected sample and does not explicitly model intrinsic absorption. Since a selection restricted to low column densities is not equivalent to a soft-band-selected sample, differences between the two should not be interpreted solely as discrepancies in the underlying LF.

Overall, our model is broadly consistent with previous X-ray determinations over the luminosity and redshift ranges where the XLFs overlap. At low redshift, the large eROSITA volume allows us to better constrain both the low- and high luminosity populations, leading to somewhat lower space densities at the bright end than in some earlier models, though the models remain within the error bars of our data. At intermediate redshifts, particularly above $z\sim1.5$, the lack of low-luminosity leverage limits our ability to constrain the faint-end slope independently. In this regime, the inferred faint-end behaviour is more sensitive to the adopted parametrisation and to the available luminosity baseline. Moreover, because the observed eROSITA band has not yet shifted fully into the rest-frame hard X-ray regime, absorption can still suppress part of the obscured population, contributing to both lower observed space densities and fluxes, relative to hard-band or absorption-corrected determinations.

\begin{figure}[t!]
    \centering
    \includegraphics[width=1\linewidth]{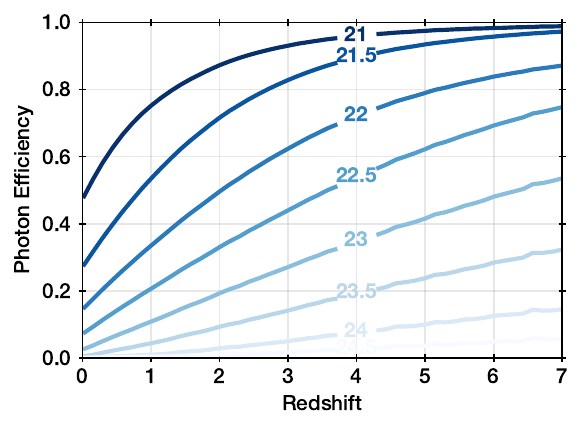}
    \caption{eROSITA main band photon detection efficiency as a function of redshift for different absorbing column densities $\log N_{\rm H}$. The observed soft X-ray band samples progressively harder rest-frame photons at higher redshifts, increasing the transmission through moderately obscuring columns.}
    \label{fig:erosens}
\end{figure}

The largest differences occur in regions that have historically been poorly sampled, though presented with small uncertainties, giving the impression that the parameter space is tightly constrained, particularly towards the bright and high redshift regimes. At $z \geq 3$, the eROSITA $0.2{-}2.3\,\mathrm{keV}$ band samples progressively harder rest-frame energies, approaching the conventional $2-10\,\mathrm{keV}$ range, thereby reducing the sensitivity of the observed flux to moderate absorption and increasing the fraction of sources with $N_{\rm H} > 10^{22}$ entering the eROSITA soft band-selected sample (see \cref{fig:erosens}). However, this advantage does not remove obscuration biases entirely, since the fraction of heavily obscured and Compton-thick AGN is observed to increase towards high redshift \citep[e.g.,][]{Aird_2015, Buchner_2015, Vito_2018, Laloux_2023, Pouliasis_2025, Ruiz_2026}. In this high-redshift regime, our fit favours a higher $L_{\star}$ and a steeper bright end in comparison to the literature XLFs in \cref{fig:collage}, consistent with \cref{fig:param_evo}. 

At high redshift, the eROSITA sample provides strong leverage on the luminous end of the XLF, but little direct information on the faint side of the break. As a result, the fitted parameters must partly compensate for this missing faint-end anchor: changes in the $C$, $L_{\star}$, and $\delta$ can produce similar predictions over the luminosity range actually sampled by the data, while implying different extrapolations towards lower luminosities. This covariance is important when comparing to literature models. For example, existing high-redshift measurements such as those of \citet{Buchner_2015, Aird_2015, Vito_2018} suggest how the LF may continue towards fainter luminosities, and a different placement of the knee would naturally shift both the inferred normalisation and characteristic break luminosity. Therefore, the differences we find relative to some previous XLF parameterisations may partly reflect the limited faint-end leverage imposed by the eROSITA sensitivity boundary and bandpass at high redshift. At the same time, they may also represent a genuine improvement in the bright-end constraints, since the large eROSITA survey volume directly probes rare luminous AGN that were only weakly sampled in earlier surveys.

\subsection{Luminosity-dependent evolution and downsizing}

\begin{figure}[t!]
    \centering
    \includegraphics[width=1\linewidth]{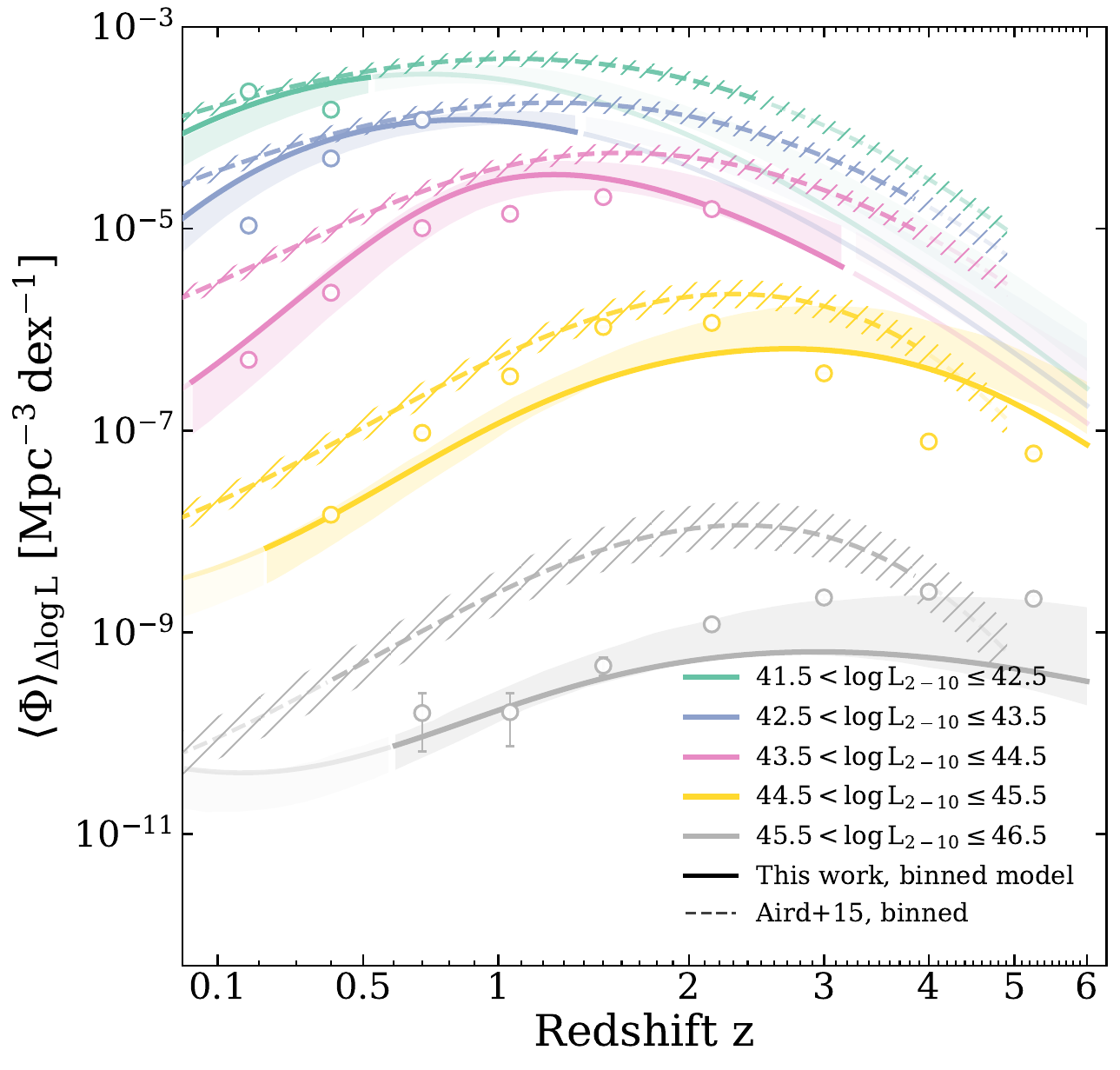}
    \caption{AGN space density evolution with redshift. The coloured solid curves show this work's best-fitting XLF and $1\sigma$ boundaries evaluated at the centres of five rest-frame $2 - 10$ keV luminosity intervals, as indicated in the legend. The high-opacity portions of the curves indicate the data-supported range of luminosity bins, while the low-opacity extensions show the model outside the directly sampled redshift intervals. Open circles show the corresponding bin-averaged \cite{Page_2000} estimates from the combined DR2 and eFEDS sample. Dashed curves show the \cite{Aird_2015}  soft-band-selected FDPL model evaluated at the same luminosities.}
    \label{fig:down}
\end{figure}

In \cref{fig:down}, we show the redshift evolution of the luminosity density inferred from the best-fitting XLF, evaluated in fixed rest-frame $2\!-\!10\,\mathrm{keV}$ luminosity intervals and compared to the corresponding \citet{Page_2000} V$_{\rm max}$ estimates. The $\Phi$-evolution is clearly luminosity dependent: moderate-luminosity AGN reach their maximum space density at lower redshift, whereas the most luminous sources peak earlier and remain comparatively important out to higher redshift. This behaviour is consistent with the established AGN downsizing picture, in which the dominant accreting population shifts from luminous quasars at earlier cosmic times to lower-luminosity AGN at later epochs \citep[e.g.,][]{Hirschmann_2012, Miyaji_2015}. 

At low luminosities, the curves decline more rapidly at high redshift than the corresponding luminosity evolution from \citet{Aird_2015}. This should not be interpreted as a direct measurement of the intrinsic high-redshift faint-end decline, but rather as reflecting the diminishing observational capability to detect low luminosity AGN at substantial redshift, due to instrumental limitations. Although the likelihood explicitly includes the eROSITA selection function, the data provide little leverage once a luminosity bin falls close to or below the survey sensitivity limit. In this regime, the effective survey volume becomes very small, and the inferred evolution is increasingly driven by the adopted parametric form and its covariance with the better-constrained regions of the XLF. The fading of the curves in \cref{fig:down}, therefore, marks the redshift coverage where the model is being extrapolated beyond the luminosity-redshift range directly populated by the data, rather than where the survey has measured a physical downturn in the AGN space density. Conversely, the highest-luminosity bins remain populated to much larger redshifts, allowing the bright end to be constrained more directly than in previous studies based on smaller survey areas.

\begin{figure*}[t!]
    
    \centering
    \includegraphics[width=1\linewidth]{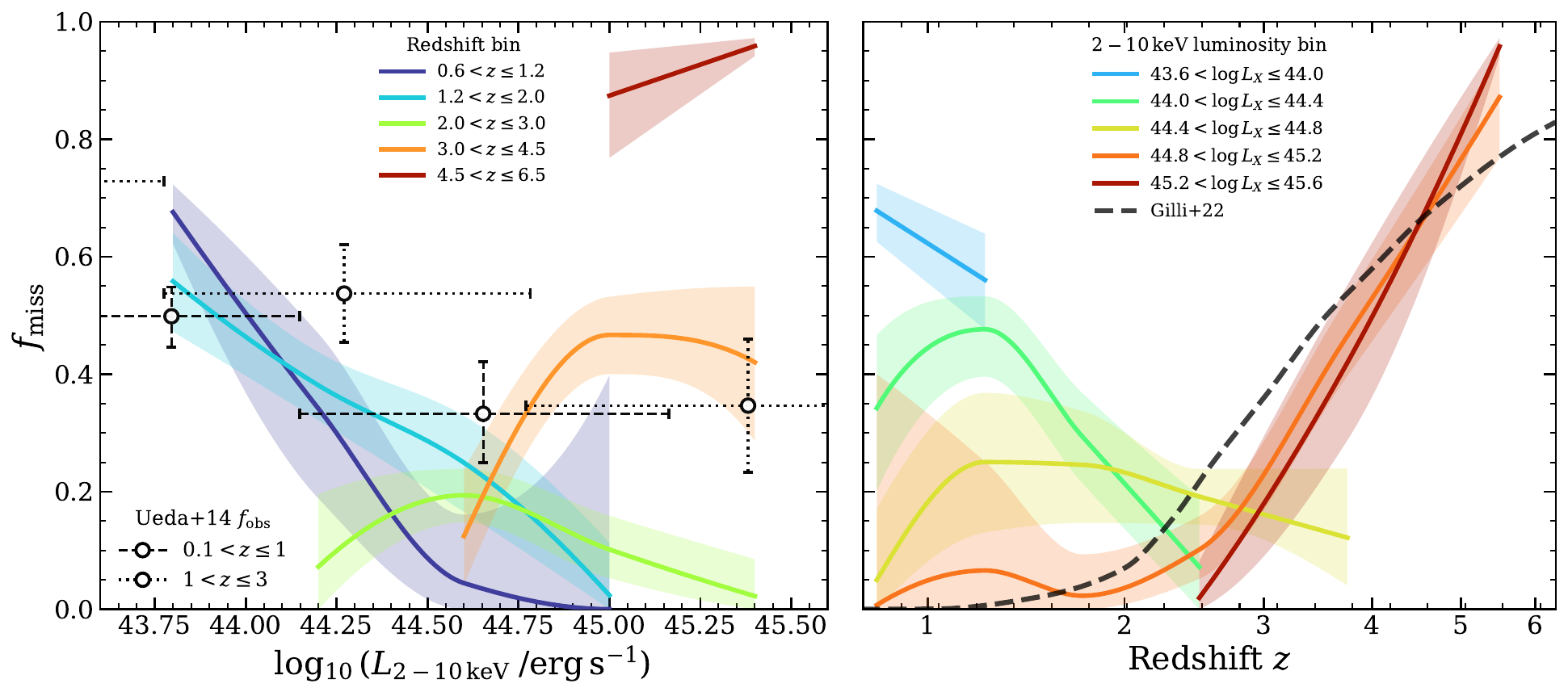}
    \caption{UV-missed fraction inferred from the comparison between optical/UV QLFs converted to rest-frame $2 - 10$ keV luminosities and this work's XLF. \emph{Left}: UV-missed fraction as a function of luminosity binned by redshift. Additionally, we plot the ISM-obscured AGN fraction ($\rm{log}\, N_{H} > 23, \, \gamma = 2$) from \cite{Gilli_2022}. \emph{Right}: UV-missed fraction as a function of redshift binned by luminosity. As a reference, we also plot the X-ray inferred obscured fraction from \cite{Ueda_2014}. Solid curves show the median in each bin, and shaded regions indicate the corresponding 1$\sigma$ interval, including the propagated uncertainty from the XLF posterior and the UV/optical LF measurements.}
    \label{fig:UV_missed_fraction}
\end{figure*}

While the overall amplitudes of \citet{Aird_2015} are broadly similar, several systematic differences are apparent. Most notably, at low redshift, our inferred space densities are lower. This is likely a consequence of the much larger eROSITA survey volume, which provides stronger constraints on all luminosity classes. These indicate lower space densities for sources with $L_{2-10\, {\rm keV}} \geq 10^{43}\,{\rm erg\,s^{-1}}$ in addition to reducing the need to extrapolate the XLF derived from smaller samples, such as those obtained from ROSAT (see \cref{fig:area_depth}) and included in \cite{Aird_2015}. At intermediate redshifts, the limited low-luminosity reach of eROSITA weakens the direct constraints on the faint-end evolution, so differences in the low-$L_X$ curves should be interpreted cautiously. At the highest redshifts, however, the bright end is better sampled, with our model predicting a shallower decline for the most luminous AGN. This suggests that previous parameterisations may underpredict the abundance of rare, luminous quasars at $z\gtrsim5$. Since our baseline fit does not explicitly correct for heavily obscured or CTK sources, any missing luminous obscured population would only increase the intrinsic space density in this regime, strengthening this conclusion. The precise amplitude of the effect, however, remains sensitive to the adopted functional form and assumed absorption correction.

\subsection{Evolution of the optical/UV-missed AGN fraction}
\label{sec:UV_missed}

UV-selected quasar LFs predominantly trace unobscured, accretion-disc-dominated systems, whereas X-ray selection is sensitive to a broader AGN population, including sources with moderate levels of obscuration and low luminosity AGN, which are outshone by their host galaxies in the optical. Comparing optical/UV Quasar LFs (QLFs) to the XLF therefore provides an empirical way to assess how much of the X-ray-selected AGN population is absent from optical/UV-selected samples. This comparison does not necessarily trace the obscured AGN fraction. Instead, it measures a UV-missed fraction, which can arise from dust extinction, UV-weak quasars, type-2 systems, and host-dominated AGN, which can all contribute to incomplete overlap between X-ray and optical/UV selection \citep[e.g.,][]{Menzel_2016, LaMassa_2024}.

We convert literature UV/optical QLF measurements from \citet{Masters_2012},
\citet{Palanque-Delabrouille_2013}, \citet{Ross_2013}, 
\citet{Manti_2017},
\citet{Matsuoka_2018}, and \citet{Harikane_2022} to rest-frame $2\!-\!10\,\mathrm{keV}$ luminosities using a common set of assumptions. Starting from absolute magnitudes at $1450\,\text{\AA}$, we first convert $M_{1450}$ to a monochromatic luminosity, $L_{\nu}(1450\,\text{\AA})$, using the AB magnitude definition. We then assume a UV continuum slope $L_{\nu}\propto\nu^{-0.5}$ \citep{Vanden_Berk_2001} to extrapolate from $1450\,\text{\AA}$ to $2500\,\text{\AA}$ and obtain $L_{2500}$ to infer the monochromatic luminosity at 2 keV from the \cite{Lusso_2016} best-fit $\alpha_{\rm ox}-L_{2500}$ relation, while accounting for the scatter. Finally, assuming an X-ray power law with photon index $\Gamma=2$, we convert the 2 keV monochromatic luminosity into an integrated rest-frame $2\!-\!10\,\mathrm{keV}$ luminosity \citep{Liu_2022}. The QLF space densities are then transformed from $\mathrm{mag^{-1}}$ to $\mathrm{dex^{-1}}$ using the corresponding Jacobian. 

The converted QLFs, $\Phi_{\rm UV\rightarrow X}$, are then compared to our best-fitting XLF, $\Phi_{\rm XLF}$, evaluated at the same luminosity and redshift. Importantly, to avoid comparing extrapolated XLF predictions with binned QLF constraints, we restrict the comparison to luminosity and redshift intervals that are directly supported by both datasets, where the fitted model is anchored by the underlying observations. We define the UV-missed fraction as
\begin{equation}
    f_{\rm UVmiss}
    =
    1 -
    \frac{\Phi_{\rm UV\rightarrow X}}
         {\Phi_{\rm XLF}}\,,
\end{equation}
where a value of $f_{\rm UVmiss}=0$ indicates that the converted UV QLF and the XLF predict the same space density, while $f_{\rm UVmiss}\rightarrow1$ implies that the UV-selected QLF recovers only a small fraction of the soft X-ray-selected population.

The left panel of \cref{fig:UV_missed_fraction} shows that at low and intermediate redshifts ($z \lesssim 3$), UV-selected QLFs recover the X-ray space density most efficiently at the highest luminosities (log$(L_{X}) \gtrsim 44.5$), where broad-line quasars dominate, and optical/UV selection is expected to be most complete. This finding is consistent with the decrease in the fraction of obscured AGN with increasing luminosity. Towards lower luminosities (log$(L_{X}) \lesssim 44.5$), the UV-missed fraction increases, reaching $\approx 0.5$ for ($z \lesssim 2$). This trend is consistent with \cite{Merloni_2014} and can be explained by a growing contribution from systems in which the accretion disc is diluted by the host galaxy or intrinsically less quasar-like in its spectral energy distribution (SED). In this regime, relatively weak X-ray emission can still identify AGN activity even when the optical/UV signatures are not sufficient for inclusion in photometric or spectroscopic QLF samples.

At higher redshift, as discussed in \cref{sec:comp_to_lit}, the observed eROSITA $0.2\!-\!2.3\,\mathrm{keV}$ band corresponds approximately to rest-frame hard X-ray energies. The comparison with UV-selected QLFs, therefore, becomes especially sensitive to AGN that are visible in (hard) X-rays but absent from optical/UV quasar samples. More specifically, eROSITA is more than 50\% complete up to $\log N_{\mathrm{H}} \leq 21,(22,23)$ at $z \geq 0,(2,6)$ (see \cref{fig:erosens}). These completeness limits were estimated by simulating large populations of AGN with physically motivated X-ray spectra over a range of intrinsic luminosities, redshifts, and absorbing column densities. The simulated spectra were folded through the eROSITA instrumental response and survey sensitivity, allowing the recovery fraction as a function of $N_{\mathrm{H}}$ and redshift to be quantified. Consistent with the V-shape predicted by \cite{Comparat_2019}, we find that the UV-missed fraction in the highest-luminosity bins (log$L_{X} \gtrsim 45$) rises sharply with redshift $z \gtrsim 2.5$, exceeding  $f_{\rm UVmiss}\sim80\%$, indicating that UV-selected surveys recover an increasingly incomplete subset of the accreting SMBH population. This behaviour is qualitatively consistent with literature and scenarios in which high-redshift AGN are embedded in gas-rich host galaxies, where obscuration may be driven not only by the nuclear torus but also by the interstellar medium of the host \citep[e.g.,][]{Liu_2017, Buchner_2017, Andonie_2024}. The obscured fractions inferred by \citet{Gilli_2022} for sources with log $N_{\rm H}\geq 23$, attributed to absorption by massive host-galaxy gas reservoirs, provide a useful comparison and in line with the high-$z$ UV-missed fractions inferred in this work.

Taken together, these results imply that optical/UV QLFs provide a good census of the most luminous unobscured quasars, but miss a substantial and luminosity-dependent fraction of the broader X-ray-selected AGN population. This missing fraction increases towards lower luminosities and appears to become particularly important at redshifts greater than those required to observe intrinsic hard X-ray emission. The result, therefore, supports the notion that a complete census of SMBH growth would require combining surveys with multiwavelength selection, including tracers less biased against obscuration.

\section{Discussion}
\label{sec5}

The XLF provides a direct census of accreting SMBHs as a function of luminosity and redshift. To translate this demographic information into a cosmic growth rate, we compute the black-hole accretion-rate density (BHAD), $\dot{\rho}_{\rm BH}(z)$. This quantity represents the rate at which mass is added to the SMBH population per unit comoving volume and time, and can be viewed as the differential form of the Soltan argument, connecting the mass contained in the predominantly dormant SMBH population observed in the local Universe to the integrated radiative energy released during earlier phases of AGN activity \citep{Soltan_1982}. Converting the XLF to a bolometric luminosity density and subsequently to a mass-accretion rate therefore links the observed AGN population to the cosmic assembly history of SMBHs. The BHAD therefore provides a physically intuitive way to simultaneously compare our XLF with previous X-ray and multiwavelength estimates of (obscured) accretion, the cosmic star-formation history, and predictions from numerical and semi-analytical models \citep[e.g.,][]{Peca_2023, Inayoshi_2024, Pouliasis_2025}.

\begin{figure*}
    \centering
    \includegraphics[width=1\linewidth]{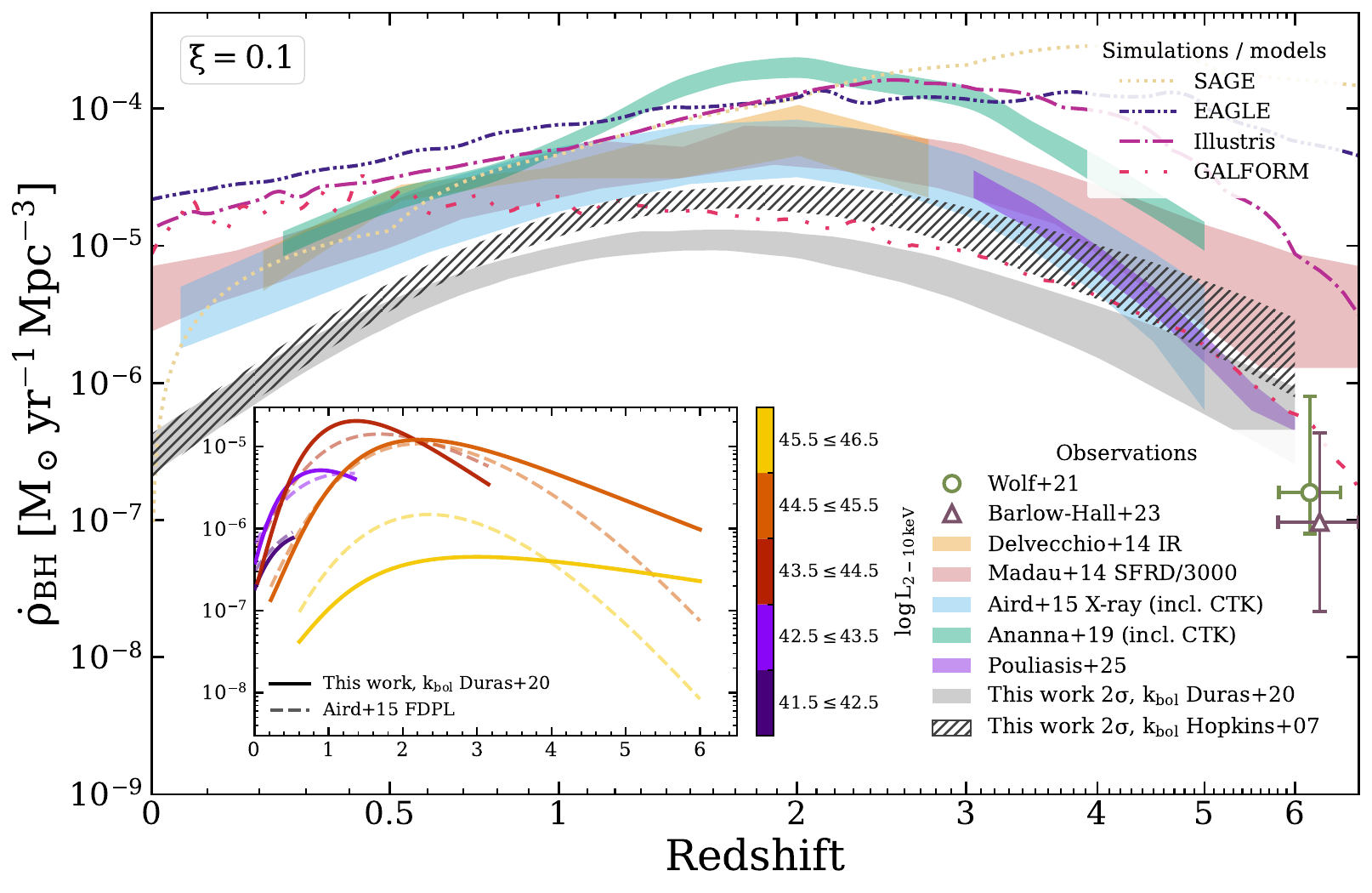}
    \caption{Cosmic BHAD from \cref{eq:BHAD} assuming  \cite{Duras_2020} \citep{Hopkins_2007} bolometric correction and a radiative efficiency of \(\xi=0.1\). The grey shaded (hatched) region indicates the bootstrap 2$\sigma$ uncertainty. Literature estimates from X-ray \citep{Aird_2015, Ananna_2019, Wolf_2021, BH_2023, Pouliasis_2025} and IR-selected \citep{Delvecchio_2014} AGN samples, the scaled cosmic star-formation-rate density \citep{Madau2014}, as well as BHAD histories inferred from simulations and semi-analytical models \citep[EAGLE, Illustris, SAGE, and GALFORM,][]{Crain_2015, Nelson_2015, Croton_2016, Griffin_2019}, are shown for comparison. The inset decomposes this work's BHAD into broad \(2\!-\!10\,\mathrm{keV}\) luminosity bins.}
    \label{fig:BHAD}
\end{figure*}

\subsection{Differential SMBH growth history}
\label{sec:BHAD}

The luminosity generated by an AGN traces only part of its total radiative output when measured in a finite X-ray band. To estimate the energy released by accretion, we convert the rest-frame $2\!-\!10\,\mathrm{keV}$ luminosity to bolometric luminosity using the luminosity-dependent correction of \citet{Duras_2020},

\begin{equation}
    L_{\rm bol} = k_{\rm bol}(L_X)\,L_X\,,
\end{equation}
where $k_{\rm bol}$ is the bolometric correction factor. The corresponding SMBH mass-growth rate follows from the conversion of accreted rest-mass energy into radiation. For a radiative efficiency $\xi$,
\begin{equation}
    L_{\rm bol}
    =
    \xi\,\dot{M}_{\rm acc}c^2\,,
\end{equation}
allowing the resulting growth rate to be expressed as
\begin{equation}
    \dot{M}_{\rm BH}(L_X)
    =
    \frac{1-\xi}{\xi}
    \frac{L_{\rm bol}(L_X)}{c^2}\,,
\end{equation}
where $c$ is the speed of light. The BHAD is consequently obtained by integrating the luminosity-dependent growth rate over the population,
\begin{equation}
    \dot{\rho}_{\rm BH}(z)
    =
    \int
    \dot{M}_{\rm BH}(L_X)\,
    \Phi(L_X,z)\,
    {\rm d}\log L_X\,.
\label{eq:BHAD}
\end{equation}
Throughout this section, we adopt $\xi=0.1$ and integrate over $40<\log L_{2-10\,\mathrm{keV}}\leq47$, unless stated otherwise. For a constant radiative efficiency, changing $\xi$ primarily rescales the normalisation of the inferred BHAD through the factor $(1-\xi)/\xi$, without altering its redshift dependence. An evolving efficiency, by contrast, could also modify the shape of the inferred growth history.

The bolometric correction introduces an additional luminosity-dependent systematic, particularly at the bright end, where $k_{\rm bol}$ can become large. To illustrate its influence, we repeat the calculation using the \citet{Hopkins_2007} correction adopted by \citet{Aird_2015}. The corresponding BHAD histories and their bootstrap $2\sigma$ ranges are shown in \cref{fig:BHAD}. Their differences, both in terms of normalisation and peak redshift, demonstrate that offsets between published BHAD estimates need not arise from the underlying XLF alone, but can also reflect the conversion from band-limited to bolometric luminosity.

We emphasise that the comparison with \citet{Aird_2015} in \cref{fig:BHAD} is intentionally different from the XLF comparisons in \cref{fig:param_evo,fig:down}. The latter use the soft-band-selected FDPL model of \citet{Aird_2015}, which provides the closest comparison to our soft-selected XLF. By contrast, \cref{fig:BHAD} shows their absorption-dependent XLAF model, including CTK AGN, to illustrate the difference between the growth history recovered from our soft-X-ray-selected population and one intended to represent a broader intrinsic AGN census. This is therefore not a like-for-like consistency test. The offset cannot be attributed uniquely to absorption as the models differ in their XLF parameterisation, bolometric treatment, and other assumptions.

The redshift of the BHAD maximum is determined by the luminosity-weighted integral of the complete XLF rather than by the evolution of any single luminosity class. As shown by the inset of \cref{fig:BHAD}, the dominant contribution in our model comes from moderate-luminosity AGN around $\log L_{2-10\,\mathrm{keV}}\sim44$, close to the knee of the XLF. These sources reach their largest contribution at $z\simeq1\!-\!2$, resulting in a total BHAD that peaks at $z\sim1.6/1.9$ when adopting the \citet{Duras_2020}/\citet{Hopkins_2007} correction. The most luminous quasars remain important at higher redshift, but their lower space densities prevent them from dominating the integrated mass-growth budget or shifting its maximum to $z\simeq2$. This behaviour can be understood more directly from the luminosity-resolved contribution to the BHAD, $\dot{M}_{\rm BH}(L_X)\Phi(L_X,z)$, shown in \cref{sec:BHAD_app}. Below $z\simeq2.5$, the difference relative to \citet{Aird_2015} is not confined to a narrow luminosity interval, but appears as an overall reduction in the luminosity-weighted space density, with the largest offset occurring at the bright end. Since these luminosities contribute disproportionately to the BHAD integral through the factor $\dot{M}_{\rm BH}(L_X)\Phi(L_X,z)$, even a modest deficit in the number density of bright AGN can shift both the normalisation and the redshift of the BHAD peak. This raises the question of how much of the low-redshift offset reflects genuine differences in the inferred XLF shape, and how much is caused by obscuration-dependent selection effects in the observed eROSITA soft band. Part of the discrepancy likely reflects differences in the direct bright-end constraints and in the adopted redshift evolution. 

Part of this difference reflects the distinct demographic leverage of the underlying surveys. The large eROSITA area directly constrains rare luminous AGN over a substantially larger volume, whereas earlier multi-tier combinations relied on smaller bright-end samples and stronger covariance among the XLF normalisation, break luminosity, and bright-end slope. The \citet{Aird_2015} model can consequently maintain a higher bright-end contribution in regimes that were more weakly sampled, while the present eROSITA measurement favours a lower luminosity-weighted space density below $z\simeq2.5$. Simultaneously, at higher redshifts, the dominant contribution shifts to increasingly luminous AGN, where bright but obscured AGN can be significantly suppressed or missed, particularly at column densities approaching the heavily obscured and Compton-thick regimes (see \cref{fig:erosens}). The offset between BHAD peaks, therefore, predominantly reflects both differences in how accretion power is distributed across the knee and bright end of the XLF, and differences in the obscured AGN fraction accounted for in each model. In particular, decompositions by column density of \cite{Buchner_2015} show that less-obscured populations tend to peak at lower redshift than more heavily obscured contributions. Additionally, several comparison curves include corrections for or remain sensitive to obscured or Compton-thick accretion, which are suppressed or absent in our soft-band-selected X-ray sample, as we do not explicitly model the full intrinsic column-density distribution.

We also compare our BHAD to estimates that include, or are less biased against obscured accretion, such as the \cite{Aird_2015} XLAF\footnote{including hard-band selected and absorption corrected sources in addition to accounting for CTK AGN} model, the CXB synthesis results of \citet{Ananna_2019}, and the IR-based estimate of \citet{Delvecchio_2014}. At high redshift, our BHAD remains consistent with the observational constraints from \citet{BH_2023}, \citet{Wolf2023}, and \citet{Pouliasis_2025} within the quoted uncertainties. This agreement is encouraging given that, at $z\gtrsim3$, the observed eROSITA $0.2{-}2.3\,\mathrm{keV}$ band corresponds approximately to rest-frame hard X-ray energies, reducing the impact of moderate obscuration, with the high-redshift BHAD being sensitive to the bright-end XLF shape and to the treatment of rare luminous AGN. When the \citet{Hopkins_2007} bolometric correction is used, our BHAD normalisation increases and becomes more similar to the XLAF-based estimate of \citet{Aird_2015}. This agreement should not be interpreted as evidence that the soft-band-selected eROSITA XLF recovers the full intrinsic accretion budget. Instead, it shows that the integrated BHAD is also sensitive to the adopted bolometric correction. Moreover, we show the cosmic star-formation-rate density (SFRD), scaled by a constant factor \citep{Madau2014}. This curve is not an independent BHAD measurement, but provides a visual reference for the relative timing of galaxy and black-hole growth. The expected scaling between SFRD and BHAD is uncertain at the factor-of-few level, depending on the assumed stellar mass return fraction, black-hole-to-stellar mass ratio,  and radiative efficiency. We therefore use it only as a guide to the relative redshift evolution. The broad similarity in shape between the scaled SFRD and the BHAD supports the established picture in which SMBH growth and galaxy growth are linked over cosmic time, although the detailed peak redshift and normalisation depend on selection and obscuration corrections. 

Finally, we compare our BHAD to those inferred from evolutionary models, including Semi-Analytic Galaxy Evolution \citep[SAGE,][]{Croton_2016}, GALFORM \citep{Griffin_2019}, Evolution and Assembly of GaLaxies and their Environments \citep[EAGLE,][]{Crain_2015}, and Illustris \citep{Nelson_2015}. However, these 
depict the intrinsic SMBH accretion histories. In contrast to X-ray population-synthesis models, they generally do not include a self-consistent line-of-sight obscuration selection or a full $N_{\rm H}$ distribution. Differences relative to the observational BHAD estimates therefore reflect that observation-based curves may be incomplete for obscured and CTK growth. The comparison highlights the large spread among theoretical predictions, consistent with previous work showing that SMBH growth histories are highly sensitive to assumptions about BH seeding, accretion, and feedback prescriptions \citep[e.g.,][]{Habouzit_2022, Habouzit_2025}. Models tend to overpredict the normalisation at both low and high redshift, although it has been argued that this offset can be attributed to observational constraints including luminosity dependence or the undersampling of sources with lower stellar mass, $M_{\star}\lesssim 10^{10}\, M_{\odot}$ \citep[e.g.][]{Habouzit_2022, Pouliasis_2025}.

\subsection{Cumulative SMBH growth history}
\label{sec:BHMD}

\begin{figure}[t!]
    \centering
    \includegraphics[width=1\linewidth]{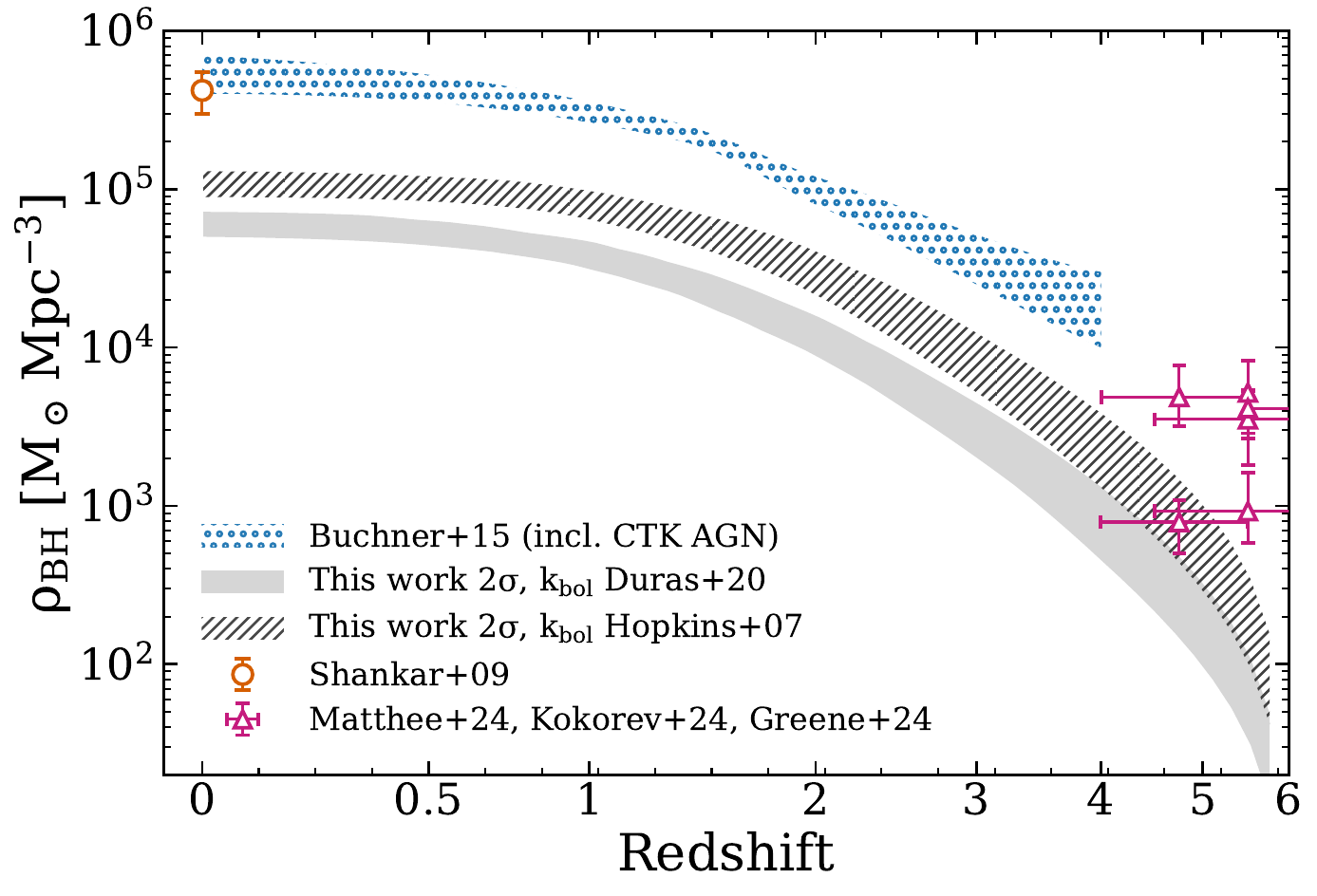}
    \caption{Cumulative BHMD inferred by integrating the best-fitting BHAD from $z=6$ to $z=0$. The grey shaded (hatched) region shows the fiducial best fit and 2$\sigma$ from this work. The bracket indicates the offset between our $z \simeq 0$ cumulative BHMD and the local value from \cite{Shankar_2009}. Pink triangles show BHMD estimates for little red dots (LRDs) taken from \cite{Matthee_2024, Greene_2024, Kokorev_2024} with the blue hatched area corresponding to the \cite{Buchner_2015} model including CTK sources.}
    \label{fig:BHMD}
\end{figure}

The BHAD describes the instantaneous growth rate of the SMBH population at a given redshift. A complementary quantity is the cumulative black-hole mass density (BHMD), $\rho_{\rm BH}(z)$, obtained by integrating the accretion-rate density over cosmic time up to the epoch denoted by $z$. This provides a direct connection to the Soltan argument and allows us to compare the amount of SMBH mass assembled through the AGN population traced by our XLF to independent estimates of the local BHMD. We compute the cumulative BHMD from the BHAD as
\begin{equation}
    \rho_{\rm BH}(z)
    =
    \int_{z_{\rm max}}^{z}
    \dot{\rho}_{\rm BH}(z')
    \frac{{\rm d}t}{{\rm d}z'}
    {\rm d}z',
\end{equation}
where
\begin{equation}
    \frac{{\rm d}t}{{\rm d}z}
    =
    \frac{1}{(1+z)H(z)} .
\end{equation}
Here, $z_{\rm max}=6$ denotes the highest redshift to which we evaluate the XLF. We therefore set $\rho_{\rm BH}(z_{\rm max})=0$, such that the resulting curve represents the mass density accumulated by radiatively efficient accretion from $z\simeq6$ to lower redshift, excluding newly seeded mass or accretion at earlier epochs, which are unlikely to contribute much. The resulting cumulative BHMD is shown in \cref{fig:BHMD}.

As expected, the cumulative mass density increases monotonically towards low redshift as the accreted mass is accumulated over cosmic time. The normalisation, however, depends noticeably on the adopted bolometric correction. Using the \citet{Duras_2020} correction, we obtain
$\rho_{\rm BH}(z\simeq0)=6.1^{+1.1}_{-1.1}\times10^{4}\,M_\odot\,{\rm Mpc^{-3}}$, while adopting the \citet{Hopkins_2007} correction increases this to
$\rho_{\rm BH}(z\simeq0)=1.10^{+0.20}_{-0.21}\times10^{5}\,M_\odot\,{\rm Mpc^{-3}}$. Both values remain below local BHMD estimates of approximately $3{-}5.5\times10^{5}\,M_\odot\,{\rm Mpc^{-3}}$ from \citet{Shankar_2009}, although those estimates are themselves affected by uncertainties in galaxy scaling relations, bulge decompositions, and SMBH occupation fractions. If this difference can be attributed entirely to accretion missed by our baseline soft X-ray-selected XLF, it would imply that only about 
$\sim14^{+10}_{-5}\%$ using the \citet{Duras_2020} correction and
$\sim26^{+17}_{-10}\%$ using the \citet{Hopkins_2007} correction are recovered. Equivalently, this implies missing fractions of
$\sim86^{+5}_{-10}\%$ and
$\sim74^{+10}_{-17}\%$, respectively. This missing component may be associated with heavily obscured and Compton-thick accretion or sources outside the adopted luminosity range. The result is broadly consistent with studies that infer a substantial obscured and Compton-thick contribution to SMBH growth \citep[e.g.][]{Buchner_2015, Ananna_2019, Comparat_2019, Kammoun_2020, Boorman_2025, Annuar_2025}.

For comparison, we also show recent estimates for little red dots (LRDs) at $z\sim5{-}6$. These studies infer bolometric LFs for LRDs and convert them to BH growth assuming $\xi=0.1$. Their respective cumulative mass density appears to exceed the BHMD inferred from this work's XLF-based growth histories at similar redshifts, but appears more consistent with models accounting for CTK sources \citep{Buchner_2015, Inayoshi_2024}. However, we note that this potential excess depends on the assumed AGN nature of the LRD population, completeness corrections, the adopted radiative efficiency, and, as discussed above, the bolometric correction.

\subsection{Future mission prospects}
\label{sec:future_prospects}

The results presented in this work highlight both the statistical power and the current limitations of wide-area X-ray AGN surveys. In practice, the limiting steps are often no longer the X-ray detections themselves, but the identification of reliable counterparts, the availability of redshift information, and the ability to model both the selection function and incompleteness. Future photometric X-ray, optical, near-infrared, as well as spectroscopic facilities will therefore improve AGN demographic studies in complementary ways. In this context, forthcoming and ongoing surveys can be viewed as strengthening different parts of the observational chain from source detection to XLF inference. Below, we briefly discuss the prospects offered by \textit{Euclid} \citep{EuclidSkyOverview}, the Legacy Survey of Space and Time \citep[LSST,][]{Ivezi2019}, the 4-metre Multi-Object Spectroscopic Telescope \citep[4MOST,][]{DeJong_2019}, and NewAthena \citep{Cruise_2025}.

\subsubsection{Improved counterparts with \textit{Euclid} and LSST}
\label{sec:euclid_lsst}

Although X-ray detections from eROSITA provide the initial evidence for accretion, the counterpart identification, redshift estimate, and host-galaxy characterisation are determined primarily from ancillary imaging data. Future deep optical and near-infrared surveys will therefore improve eROSITA AGN studies not only by detecting fainter counterparts, but also by reducing the ambiguity of the associations themselves \citep{Roster_2025_euclid}.

These include \textit{Euclid}, combining high angular resolution, depth, and near-infrared coverage. This will provide lower positional uncertainties and less noisy photometry for X-ray sources that are faint, blended, or missing in current optical catalogues, leading to more reliable counterpart associations. The near-infrared coverage is especially important at $z\gtrsim2$, where key rest-frame optical features, such as the Balmer/4000\,\AA{} break, move out of the optical bands. Likewise, the Vera C. Rubin Observatory, which recently started operations, will provide a complementary optical data set through its wide-area, deep multi-band imaging and time-domain coverage. Its depth will increase the fraction of eROSITA sources with optical counterparts, while variability information will help in rejecting stellar or inactive-galaxy contaminants. Together, \textit{Euclid} and LSST will extend the SED wavelength baseline, reducing colour--redshift degeneracies and improving photometric classification.

\subsubsection{Spectroscopic follow-up with 4MOST}
\label{sec:4most}

Large spectroscopic surveys have transformed extragalactic astronomy by providing redshifts for millions of sources. DESI, for example, combines a 4-m telescope with 5000 fibres over an $\sim8\,{\rm deg^2}$ field of view and is exceptionally efficient for large northern-sky galaxy and quasar samples. SDSS-V provides complementary, flexible spectroscopy from both hemispheres using two 2.5-m telescopes, including repeated observations. However, their selection functions are primarily set by optical target selection and therefore favour comparatively bright sources, many of which are inactive galaxies. X-ray-selected AGN samples probe a different and more heterogeneous population: their counterparts can be optically faint or otherwise poorly represented in existing spectroscopic catalogues. Consequently, despite dedicated SDSS-V black-hole science programmes, photo-$z$s remain the only practical route to near-complete redshift coverage for eROSITA-scale samples. 

In this context, 4MOST will be especially important for eROSITA. With 2436 fibres over an $\sim4.1\,{\rm deg^2}$ field of view on the VISTA telescope, it can be viewed as a southern wide-field spectroscopic counterpart to DESI, but with dedicated eROSITA follow-up programmes already built into its survey strategy \citep{DeJong_2019}. Its southern footprint is particularly well matched to the eROSITA-DE sky, and its depth enables spectroscopy for many X-ray counterparts that are fainter than, or not prioritised by, existing wide-area spectroscopic catalogues \citep{Merloni_2019}. As such, 4MOST will additionally support the validation of photo-$z$s, identify catastrophic failures, and improve the calibration of future eROSITA AGN XLF measurements.

We can estimate the potential impact of 4MOST using a target catalogue based on real eRASS:4 X-ray sources with LS10 counterparts, restricted to $-80^\circ<{\rm Dec}<+5^\circ$ and optical magnitude $r<22.8$, together with simulations of the 4MOST observing process. The simulations assign each target an exposure-completion metric, $f_{\rm obs}$, which quantifies the fraction of exposure time accumulated relative to that required to reach the target spectral signal-to-noise for reliable redshift measurement. Thus, $f_{\rm obs}>0.1$ indicates that a source receives some non-negligible exposure, while $f_{\rm obs}>1$ corresponds approximately to sufficient exposure for a highly reliable redshift determination. Importantly, $f_{\rm obs}<1$ does not imply that no redshift can be measured, but rather that the expected reliability is less certain.

Applying the same footprint and optical magnitude limits to the DR2 catalogue released in this work and cross-matching its 821\,441 entries with this 4MOST target catalogue, yields approximately $6.4\times10^{5}$ common sources, roughly $42\%$ of which already carry spec-$z$ from our literature compilation of redshifts. This comparison should be interpreted with some caution, since counterpart assignments can change between eRASS data releases as the processing pipeline, X-ray positions, and source properties are updated. Of the matched sources, approximately $5.4\times10^{5}$, or $84\%$, have $f_{\rm obs}>0.1$, while $4.7\times10^{5}$, or $73\%$, have $f_{\rm obs}>1$. Correspondingly, we find approximately $3.1\times10^{5}$ sources with $f_{\rm obs}>0.1$ and $2.5\times10^{5}$ sources with $f_{\rm obs}>1$ that currently lack a known spec-$z$. These objects represent a particularly valuable subset: they are already identified as likely X-ray-selected AGN and have suitable multiwavelength counterparts. As such, we estimate that 4MOST would increase the spectroscopic completeness of the DR2 AGN sample from $\sim30\%$ to $\gtrsim60\%$.

The final number of new secure redshifts will depend on target prioritisation, fibre allocation, observing conditions, source brightness, and the achieved redshift-success rate. Even so, within the footprint and brightness limits described above, this simple estimate suggests that 4MOST could provide on the order of $(2.5{-}3.1)\times10^{5}$ new spec-$z$s for eRASS:4-selected sources currently lacking spectroscopy. Such an increase would not only improve the number of sources with secure luminosities but also address one of the main systematic limitations in future eROSITA LF measurements: the propagation of redshift incompleteness and photo-$z$ failures into inferred AGN space densities.

\subsubsection{Future X-ray constraints with NewAthena}
\label{sec:newathena}

\begin{figure}
    \centering
    \includegraphics[width=1\linewidth]{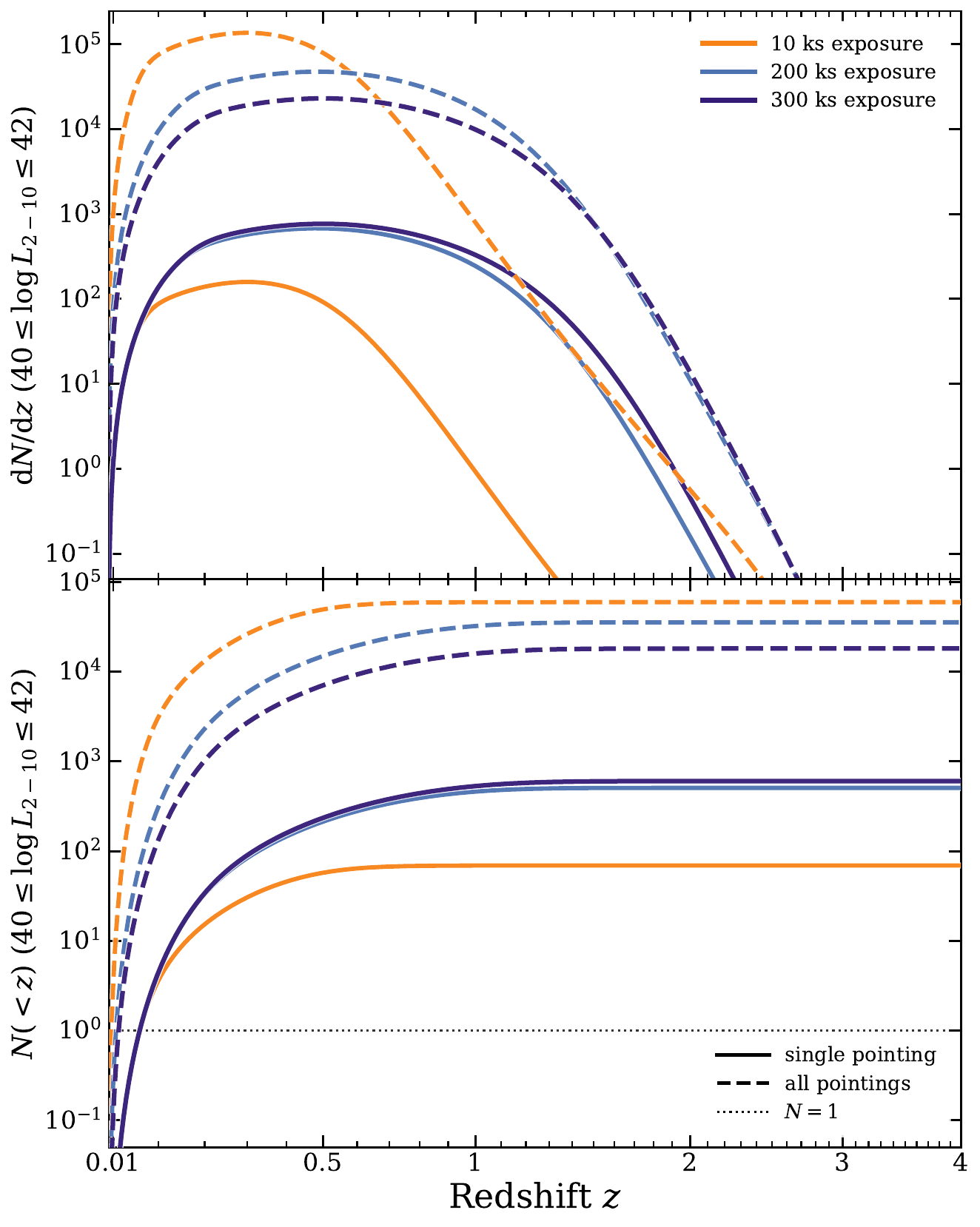}
    \caption{Expected NewAthena detections of faint AGN in the luminosity interval $40\leq\log L_{2-10\,\mathrm{keV}}\leq42$, obtained by folding the hybrid XLF model through the area curves for the 10, 200, and 300~ks survey tiers. The upper panel shows the differential yield, ${\rm d}N/{\rm d}z$, while the lower panel shows the cumulative number of sources expected below a given redshift. Solid lines correspond to a single pointing, and dashed lines show the yield after multiplying by the planned number of pointings in each tier (860, 70, and 30, respectively).}
    \label{fig:athena_curves}
\end{figure}

\begin{figure*}[t!]
    
    \centering
    \includegraphics[width=1\linewidth]{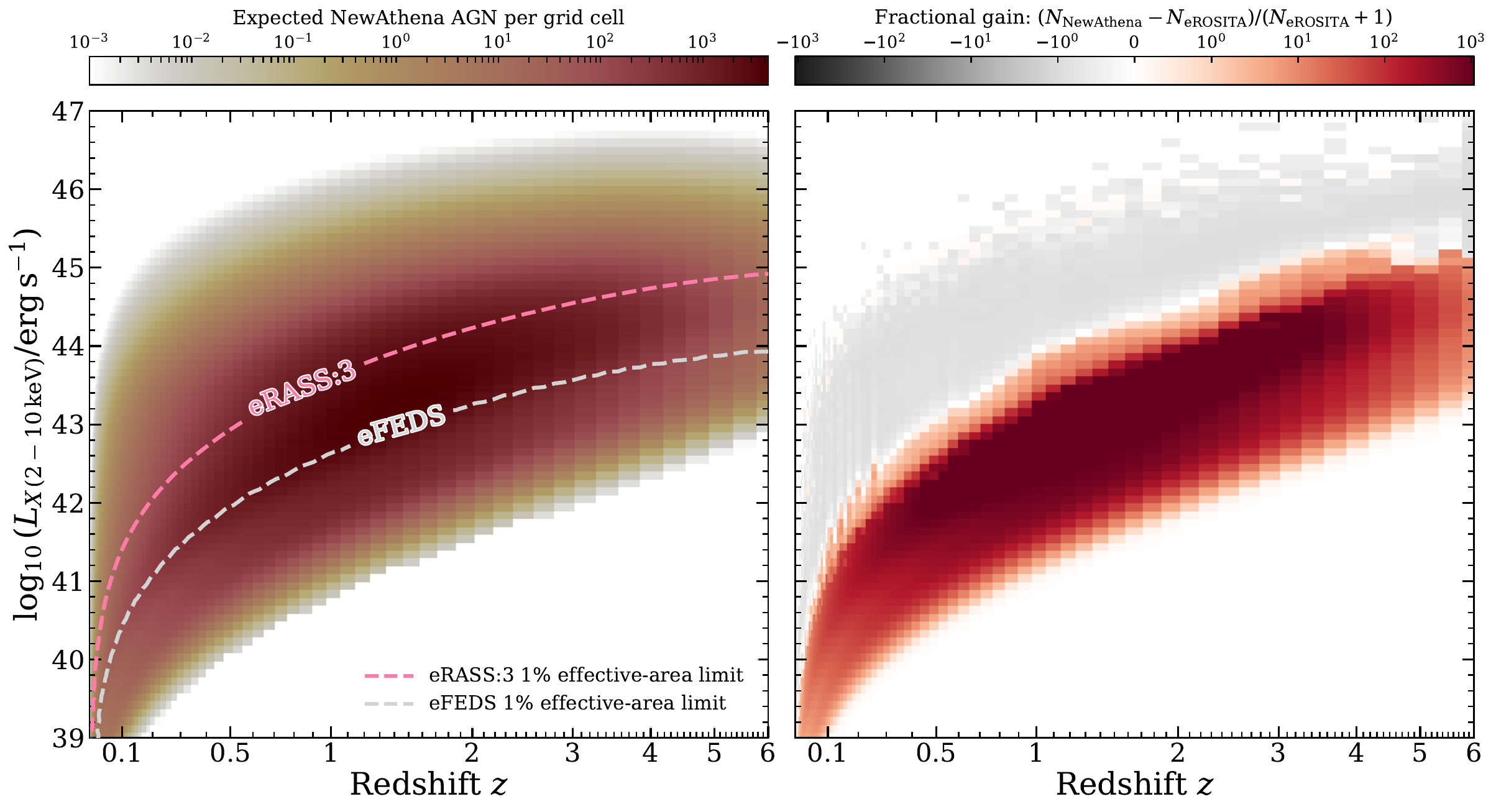}
    \caption{Forecasted NewAthena AGN yield in the luminosity--redshift plane. \emph{Left}: Expected number of AGN per grid cell for the combined NewAthena survey, obtained by folding a hybrid XLF model through the flux-dependent area curves of the 10, 200, and 300~ks survey tiers. \emph{Right}: Fractional gain relative to the XLF eRASS+eFEDS sample used in this work, defined in each luminosity--redshift cell. Dashed contours indicate the $1\%$ effective-area limits of the XLF eRASS:3 and eFEDS samples, converted to rest-frame $2\!-\!10\,\mathrm{keV}$ luminosities.}
    \label{fig:athena_comp}
\end{figure*}

Future progress on AGN demographics will also require deeper X-ray observations that extend beyond the luminosity--redshift region currently accessible to eROSITA. While eROSITA provides the large-area statistical baseline needed to constrain luminous AGN, its soft-band sensitivity limits the direct characterisation of moderate-luminosity and obscured systems at high redshift. NewAthena will address this complementary regime through its larger collecting area, broad X-ray bandpass, and substantially improved sensitivity, enabling deeper surveys over well-defined areas of the sky.

To illustrate the potential gain, we fold a hybrid XLF model through representative NewAthena survey-tier area curves. In the $L_{X}-z$ region directly constrained by eROSITA, we use the best-fitting soft-band XLF derived in this work. At lower luminosities, where the XLF in \cref{fig:collage} is no longer data supported, we splice in the soft-selected FDPL model of \citet{Aird_2015}. This choice is intended to provide an apples-to-apples forecast for the soft-X-ray-visible AGN population, rather than a fully intrinsic absorption-corrected AGN census. The predicted number of sources in a $L_{X}-z$ bin is computed as

\begin{equation}
    \begin{split}
    N_{\rm exp}
    =
    \int_{\Delta z}
    \int_{\Delta \log L_X}
    \Phi(L_X,z)\,
    A(L_X,z)\, \\
    \frac{{\rm d}V}{{\rm d}z\,{\rm d}\Omega}
    \,{\rm d}z\,{\rm d}\log L_X\, ,
    \end{split}
\end{equation}
where $A(L_X,z)$ is the flux-dependent effective area of the relevant NewAthena survey tier after converting luminosity to observed-frame flux. We consider the proposed survey components comprising 860, 70, and 30 pointings with exposures of 10, 200, and 300~kiloseconds (ks), respectively, each covering $40\times40$ arcmin$^2$. 

As a first illustration, the two panels of \cref{fig:athena_curves} show the expected number of faint AGN in the luminosity interval $40\leq\log L_{2-10\,\mathrm{keV}}\leq42$ as a function of redshift for a single pointing in each tier, as well as for the full number of planned pointings. The shape of the differential curves in the upper panel reflects the competition between survey sensitivity, cosmological volume, and XLF evolution. In the local Universe, such sources are bright and easily detectable, but the comoving volume enclosed by a redshift bin is small, so the expected number remains low. Towards intermediate redshift, the accessible volume grows rapidly while the sources remain detectable, and the XLF itself rises towards the epoch where moderate-luminosity AGN are most abundant (see \cref{fig:down}). At higher redshift, the observed flux falls below the sensitive part of the area curve and the intrinsic space density of these moderate-luminosity systems declines, causing the expected number to decrease again. In this sense, the turnover of the curves is the yield-space analogue of AGN downsizing: low and moderate-luminosity accretion becomes increasingly common up to $z\sim1$. The cumulative curves in the lower panel provide the survey-planning view: they show how many faint AGN are expected below a given redshift for a single pointing and for the full tier.

In \cref{fig:athena_comp}, we compute the expected yield for the combined NewAthena survey across the full $L_{X}-z$ plane. Here, the left panel gives the expected number of AGN, while right panel compares this forecast to the current eRASS+eFEDS XLF sample used in this work, showing the fractional gain in each cell. The gain is largest where the current X-ray census is limited by depth rather than area: below the eROSITA sensitivity boundary, at moderate luminosities and intermediate-to-high redshifts. By contrast, at the very bright end the gain is more limited, since eROSITA already samples large cosmic volumes and the number of such sources is set mainly by their intrinsic rarity. Moreover, this forecast does not include an explicit intrinsic column-density distribution or Compton-thick correction. Instead, it predicts the yield of AGN belonging to the observed soft-X-ray-visible population used to construct the baseline XLF. NewAthena's broader $0.1{-}10\,\mathrm{keV}$ bandpass and improved sensitivity will make it more capable than eROSITA of detecting obscured AGN. Including the full intrinsic obscured population would subsequently increase the expected yield, although the magnitude of this increase predominantly depends on the assumed $N_{\rm H}$ distribution and CTK fraction.
The largest scientific gain is therefore expected for the faint side of the XLF knee beyond $z\gtrsim2$. Combined with the expected advances in the optical/UV by the likes of $Euclid$ and LSST, probing this regime will further extend and constrain the nature of the UV-missed AGN population, especially towards fainter luminosities (see \cref{fig:UV_missed_fraction}). A particularly direct application will be the X-ray follow-up of high-redshift quasars identified by \textit{Euclid} and LSST \citep[e.g.,][]{Yang_2026, Belladitta_2026}. Under standard UV-to-X-ray conversions based on the $\alpha_{\rm ox}-L_{2500}$ relation, a quasar with $M_{1450}\simeq-24$ at $z\simeq7{-}8$ corresponds to an expected $2-10\,{\rm keV}$ X-ray flux of a few $10^{-16}\,\mathrm{erg\,cm^{-2}\,s^{-1}}$, depending on the assumed spectrum. Such sources are within reach of deep current facilities such as \textit{Chandra} and \textit{XMM-Newton} in targeted observations, but NewAthena would provide a more efficient route to characterising them as a population. Its larger collecting area and wide field would allow X-ray detections, or meaningful stacked constraints, for samples of optically selected high-redshift quasars, testing whether they follow standard $\alpha_{\rm ox}$ expectations or are systematically X-ray weak or obscured. A related science case is the emerging population of broad-line AGN, including LRDs, selected by the James Webb Space Telescope \citep[JWST,][]{Gardner_2006, Gardner_2023}. Several of these sources are currently undetected in recent \textit{Chandra} observations, including in stacked analyses, leaving open whether they are heavily obscured, intrinsically X-ray weak, unusually soft, or characterised by different coronal physics \citep[e.g.][Boorman et al. in prep.]{Sacchi_2025, Maiolino_2025, Comastri_2026}. While NewAthena will be particularly valuable in this context, providing deeper X-ray constraints for statistically meaningful samples, observations of the faintest and most ambiguous accretors at high-redshift may require instrumental capabilities beyond sensitivity alone. Reaching fluxes of order $10^{-18}\,\mathrm{erg\,cm^{-2}\,s^{-1}}$ will demand excellent angular resolution and low background in order to avoid source confusion. In this regime, future high-resolution X-ray mission concepts such as Lynx \citep{Gaskin_2019} or the advanced X-ray imaging satellite \citep[AXIS,][]{Reynolds_2023} would be highly complementary to NewAthena.


\section{Conclusions}
\label{sec6}

In this work, we have presented a new measurement of the soft X-ray luminosity function of AGN using the second eROSITA-DE data release, based on the stacked eRASS:3 observations of the western Galactic hemisphere, complemented by the deeper eFEDS field. A key outcome of this work is the construction and release of the eRASS DR2 AGN catalogue, including, where available, spectroscopic and otherwise photometric redshift information for the multiwavelength counterpart associations \citep{Ramos_2026}. We summarise the main results below.

\begin{itemize}
    
    \item We improve the constraints on wide-area X-ray AGN demographics significantly by providing a substantially larger and more homogeneous basis than previous multi-mission XLF compilations, including the use of full photo-$z$ PDFs (see \cref{sec2}).

    \item We introduce a redshift-dependent smoothly broken power-law XLF model in which the normalisation, break luminosity, slopes, and knee width evolve continuously (see \cref{sec3}). Compared to previous studies, we find that the break luminosity evolves more strongly with redshift, the normalisation peaks at comparatively lower redshift. In addition, we show that the bright end slope increases with redshift while the knee width is more narrow than inferred from past double-broken powerlaw XLFs. 

    \item Our XLF is broadly consistent with previous determinations where the data overlap (see \cref{sec4}). However, the large eROSITA volume favours lower bright-end space densities at low redshift, while at high redshift it provides direct leverage on rare luminous AGN, implying higher abundances than some previous extrapolations.

    \item Comparing to optical/UV QLFs converted to $2\!-\!10\,\mathrm{keV}$ luminosities, we infer the UV-missed fraction to increase at lower luminosities and, for the brightest and rarest luminous systems, again towards high redshift (see \cref{sec:UV_missed}).

    \item Integrating the XLF with a luminosity-dependent bolometric correction yields a BHAD peaking at $z\simeq1{-}1.5$, somewhat below several reference models (see \cref{sec:BHAD}). The accretion-rate density is dominated by moderate-luminosity AGN, while luminous quasars remain important at high redshift but are too rare to dominate the total growth budget.

    \item The cumulative BHMD from $z=6$ to the present remains below local BHMD estimates, with the baseline soft-band-selected XLF accounting for only $\sim14-26\%$ of the local SMBH mass density, depending on the bolometric correction (see \cref{sec:BHMD}). This is consistent with substantial SMBH growth occurring in obscured or Compton-thick phases not explicitly reconstructed in our baseline model.
     
    \item We release the eRASS DR2 AGN catalogue, comprising $\sim1.2\times10^{6}$ sources. In addition to photo-$z$ estimates for all sources, the released catalogue includes nearly $4\times10^{5}$ spec-$z$s, as a meaningful resource for follow-up studies.
    
\end{itemize}

Looking ahead, the prospects for improving AGN demographic measurements are strong. Future eROSITA releases, beginning with deeper cumulative catalogues such as eRASS:5, will substantially increase the soft X-ray sensitivity, yielding an order-of-magnitude increase in the AGN XLF sample, expanding it from currently hundreds of thousands to several million sources. Subsequently, an important future improvement will be to update the eROSITA selection function for lower-significance detections. The current analysis deliberately restricts the sample to the conservative threshold for which the area curves are defined. Extending the selection function to fainter sources will increase the sample size and improve constraints on the faint side of the XLF, particularly at redshifts where the present sample is sensitivity-limited.

At the same time, upcoming and ongoing multiwavelength surveys will address the main limitations that remain in the present analysis (see \cref{sec:future_prospects}). \textit{Euclid} and LSST will improve counterpart association and photometric redshift performance, particularly for faint and high-redshift sources. 4MOST will substantially increase the spectroscopic completeness of the eROSITA AGN sample over the southern sky. Further in the future, NewAthena will extend the X-ray census to fainter luminosities and more obscured populations, providing a foundation for the next generation of multiwavelength AGN demographic studies by probing the regions of luminosity--redshift space where the current eROSITA XLF and previous literature XLFs still require extrapolation.

\section{Data availability}

This paper presents the DR2 (main and hard) AGN catalogues, including optical counterparts from LS10 and associated redshift information, adopting spectroscopic redshifts where available and photometric redshifts otherwise. The catalogue will be publicly released via CDS and the eROSITA website\footnote{\url{https://erosita.mpe.mpg.de/dr2/AllSkySurveyData\_dr2/}}, together with the data model and column descriptions.


\begin{acknowledgements}
CA acknowledges support and resources from the Alexander von Humboldt Foundation. RJA was supported by FONDECYT grant number 1231718 and by the ANID BASAL project FB210003. This research was supported by the Excellence Cluster ORIGINS which is funded by the Deutsche Forschungsgemeinschaft (DFG, German Research Foundation) under Germany's Excellence Strategy - EXC-2094-390783311. The simulations have been carried out on the computing facilities of the Computational Center for Particle and Astrophysics (C2PAP). BT acknowledges support by the European Research Council (ERC) under the European Union's Horizon 2020 research and innovation program (grant agreement number 950533). This research was supported by the Excellence Cluster ORIGINS and by the Munich Institute for Astro-, Particle and BioPhysics (MIAPbP), which are funded by the Deutsche Forschungsgemeinschaft (DFG, German Research Foundation) under Germany's Excellence Strategy - EXC 2094 - 390783311. This work is based on data from eROSITA, the soft X-ray instrument aboard SRG, a joint Russian-German science mission supported by the Russian Space Agency (Roskosmos), in the interests of the Russian Academy of Sciences represented by its Space Research Institute (IKI), and the Deutsches Zentrum für Luft- und Raumfahrt (DLR). The SRG spacecraft was built by Lavochkin Association (NPOL) and its subcontractors and is operated by NPOL with support from the Max Planck Institute for Extraterrestrial Physics (MPE).

The development and construction of the eROSITA X-ray instrument were led by MPE, with contributions from the Dr. Karl Remeis Observatory Bamberg \& ECAP (FAU Erlangen-Nuernberg), the University of Hamburg Observatory, the Leibniz Institute for Astrophysics Potsdam (AIP), and the Institute for Astronomy and Astrophysics of the University of T\"ubingen, with the support of DLR and the Max Planck Society. The Argelander Institute for Astronomy of the University of Bonn and the Ludwig Maximilians Universit\"at Munich also participated in the science preparation for eROSITA.

The Legacy Surveys consist of three individual and complementary projects: the Dark Energy Camera Legacy Survey (DECaLS; Proposal ID 2014B-0404; PIs: David Schlegel and Arjun Dey), the Beijing-Arizona Sky Survey (BASS; NOAO Prop. ID 2015A-0801; PIs: Zhou Xu and Xiaohui Fan), and the Mayall z-band Legacy Survey (MzLS; Prop. ID 2016A-0453; PI: Arjun Dey). DECaLS, BASS, and MzLS together include data obtained, respectively, at the Blanco telescope, Cerro Tololo Inter-American Observatory, NSF’s NOIRLab; the Bok telescope, Steward Observatory, University of Arizona; and the Mayall telescope, Kitt Peak National Observatory, NOIRLab. Pipeline processing and analyses of the data were supported by NOIRLab and the Lawrence Berkeley National Laboratory (LBNL). The Legacy Surveys project is honored to be permitted to conduct astronomical research on Iolkam Du’ag (Kitt Peak), a mountain with particular significance to the Tohono O’odham Nation.

NOIRLab is operated by the Association of Universities for Research in Astronomy (AURA) under a cooperative agreement with the National Science Foundation. LBNL is managed by the Regents of the University of California under contract to the U.S. Department of Energy.

This project used data obtained with the Dark Energy Camera (DECam), which was constructed by the Dark Energy Survey (DES) collaboration. Funding for the DES Projects has been provided by the U.S. Department of Energy, the U.S. National Science Foundation, the Ministry of Science and Education of Spain, the Science and Technology Facilities Council of the United Kingdom, the Higher Education Funding Council for England, the National Center for Supercomputing Applications at the University of Illinois at Urbana-Champaign, the Kavli Institute of Cosmological Physics at the University of Chicago, Center for Cosmology and Astro-Particle Physics at the Ohio State University, the Mitchell Institute for Fundamental Physics and Astronomy at Texas A \& M University, Financiadora de Estudos e Projetos, Fundacao Carlos Chagas Filho de Amparo, Financiadora de Estudos e Projetos, Fundacao Carlos Chagas Filho de Amparo a Pesquisa do Estado do Rio de Janeiro, Conselho Nacional de Desenvolvimento Cientifico e Tecnologico and the Ministerio da Ciencia, Tecnologia e Inovacao, the Deutsche Forschungsgemeinschaft and the Collaborating Institutions in the Dark Energy Survey. The Collaborating Institutions are Argonne National Laboratory, the University of California at Santa Cruz, the University of Cambridge, Centro de Investigaciones Energeticas, Medioambientales y Tecnologicas-Madrid, the University of Chicago, University College London, the DES-Brazil Consortium, the University of Edinburgh, the Eidgenossische Technische Hochschule (ETH) Zurich, Fermi National Accelerator Laboratory, the University of Illinois at Urbana-Champaign, the Institut de Ciencies de l’Espai (IEEC/CSIC), the Institut de Fisica d’Altes Energies, Lawrence Berkeley National Laboratory, the Ludwig Maximilians Universitat Munchen and the associated Excellence Cluster Universe, the University of Michigan, NSF’s NOIRLab, the University of Nottingham, the Ohio State University, the University of Pennsylvania, the University of Portsmouth, SLAC National Accelerator Laboratory, Stanford University, the University of Sussex, and Texas A\&M University.

BASS is a key project of the Telescope Access Program (TAP), which has been funded by the National Astronomical Observatories of China, the Chinese Academy of Sciences (the Strategic Priority Research Program “The Emergence of Cosmological Structures” Grant \# XDB09000000), and the Special Fund for Astronomy from the Ministry of Finance. The BASS is also supported by the External Cooperation Program of Chinese Academy of Sciences (Grant \# 114A11KYSB20160057), and Chinese National Natural Science Foundation (Grant \# 12120101003, \# 11433005).

The Legacy Survey team uses data products from the Near-Earth Object Wide-field Infrared Survey Explorer (NEOWISE), a project of the Jet Propulsion Laboratory/California Institute of Technology. NEOWISE is funded by the National Aeronautics and Space Administration.

The Legacy Surveys imaging of the DESI footprint is supported by the Director, Office of Science, Office of High Energy Physics of the U.S. Department of Energy under Contract No. DE-AC02-05CH1123, by the National Energy Research Scientific Computing Center, a DOE Office of Science User Facility under the same contract, and by the U.S. National Science Foundation, Division of Astronomical Sciences under Contract No. AST-0950945 to NOAO.

Funding for the Sloan Digital Sky Survey V has been provided by the Alfred P. Sloan Foundation, the Heising-Simons Foundation, the National Science Foundation, and the Participating Institutions. SDSS acknowledges support and resources from the Center for High-Performance Computing at the University of Utah. SDSS telescopes are located at Apache Point Observatory, funded by the Astrophysical Research Consortium and operated by New Mexico State University, and at Las Campanas Observatory, operated by the Carnegie Institution for Science. The SDSS web site is \url{www.sdss.org}.

SDSS is managed by the Astrophysical Research Consortium for the Participating Institutions of the SDSS Collaboration, including Caltech, The Carnegie Institution for Science, Chilean National Time Allocation Committee (CNTAC) ratified researchers, The Flatiron Institute, the Gotham Participation Group, Harvard University, Heidelberg University, The Johns Hopkins University, L’Ecole polytechnique f\'{e}d\'{e}rale de Lausanne (EPFL), Leibniz-Institut f\"ur Astrophysik Potsdam (AIP), Max-Planck-Institut f\"ur Astronomie (MPIA Heidelberg), Max-Planck-Institut f\"ur Extraterrestrische Physik (MPE), Nanjing University, National Astronomical Observatories of China (NAOC), New Mexico State University, The Ohio State University, Pennsylvania State University, Smithsonian Astrophysical Observatory, Space Telescope Science Institute (STScI), the Stellar Astrophysics Participation Group, Universidad Nacional Aut\'{o}noma de M\'exico, University of Arizona, University of Colorado Boulder, University of Illinois at Urbana-Champaign, University of Toronto, University of Utah, University of Virginia, Yale University, and Yunnan University.
\end{acknowledgements}

%
\bibliographystyle{aa.bst} 
\bibliography{bibliography} 

\FloatBarrier
\begin{appendix}

\section{Photometric redshift performance}

As for eROSITA DR1 \citep[see][]{Salvato25}, the accuracy $\sigma_{\rm NMAD} = 1.4826 \times {\rm median} \, \frac{|z_{\rm spec}-z_{\rm phot}|}{(1+z_{\rm spec})}$ of DR2 is roughly 0.07, but with a slightly higher fraction of outliers $\eta = \frac{|z_{\rm spec}-z_{\rm phot}|}{(1+z_{\rm spec})} > 0.15$ of 17\% instead of 13\%, due to the increased fraction of faint ($r$ > 22) LS10 CTPs with larger uncertainties. 

\begin{figure}[htbp]
    \centering
    \includegraphics[width=1\linewidth]{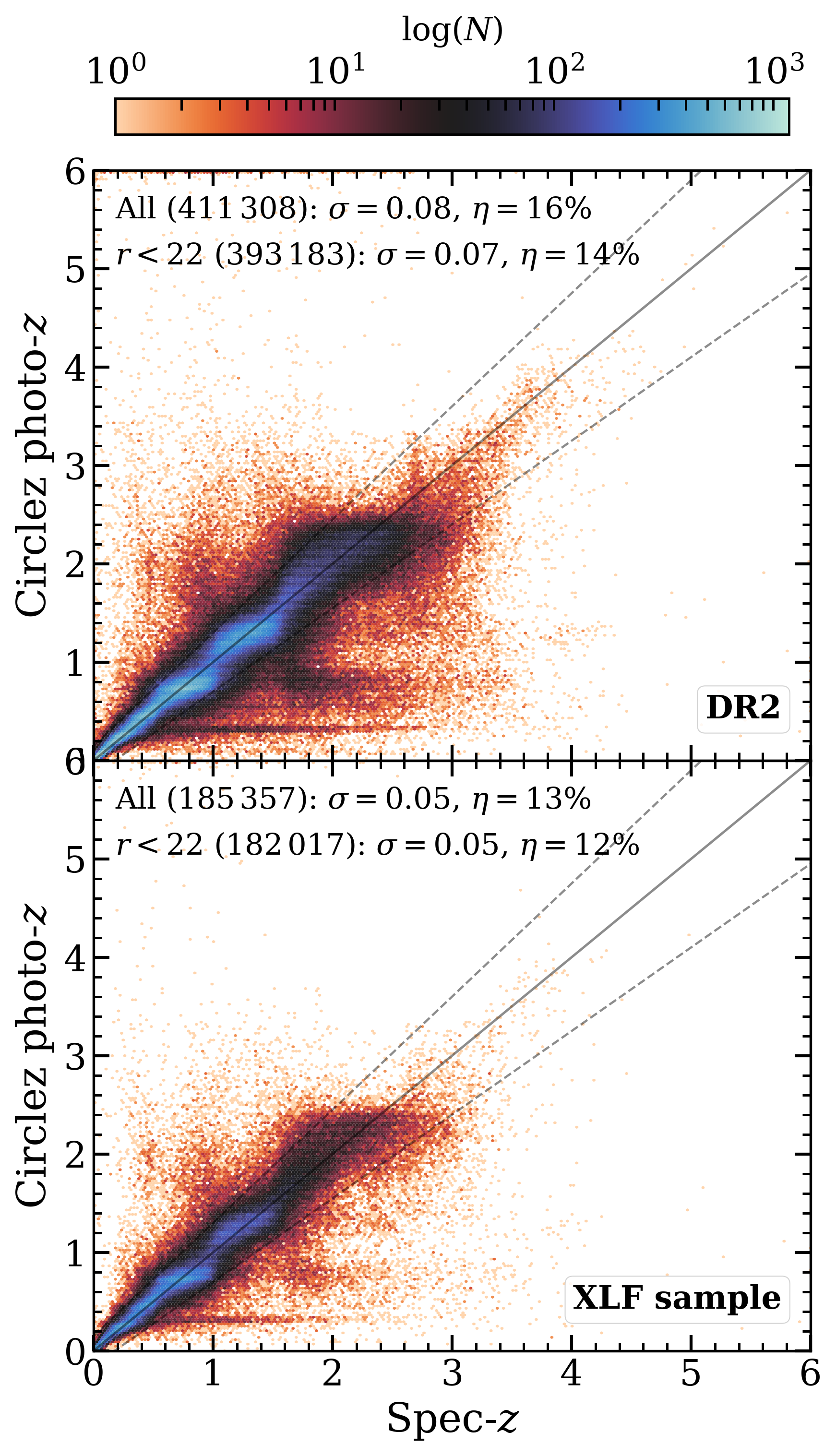}
    \caption{Circlez photo-$z$ performance for the released DR2 catalogue (top) and subselected XLF sample (bottom). Outliers, i.e., where $\frac{|\Delta z|}{1+z_{\rm spec}} > 0.15, $ correspond to sources outside the area defined by the dashed lines. The apparent horizontal banding is a visualisation artefact and does not reflect a feature of the redshift estimation method.}
    \label{fig:redshifts}
\end{figure}

\section{Model-independent space density}
\label{sec:vmax}

As an independent consistency check of the forward-modelled XLF, we compute binned LF estimates using the estimator introduced by \citet{Page_2000}. Unlike the original $1/V_{\max}$ approach of \citep{Schmidt_1968}, this method estimates the space density in a luminosity-redshift bin from the number of detected sources divided by the accessible survey volume integrated across the bin. For a given X-ray luminosity bin $[L_{\rm min}, L_{\rm max}]$ and redshift interval $[z_{\rm min}, z_{\rm max}(L)]$, the luminosity function is estimated as
\begin{equation}
\frac{\mathrm{d}\phi(L,z)}{\mathrm{d}\,{\rm log}L} = \frac{n}{\int_{{\rm log}L_{\rm min}}^{{\rm log}L_{\rm max}} \int_{z_{\rm min}}^{z_{\rm max}(L)} A(L, z)\, \frac{\mathrm{d}V}{\mathrm{d}z}\,\mathrm{d}z\,\mathrm{d}\,{\rm log}L}\, ,
\end{equation}
where $n$ is the number of sources in the bin. Agreement between the binned $1/V_{\max}$ estimates and the best-fit model, therefore, serves as a qualitative validation of the inferred XLF.

\section{Literature XLF recovery}
\label{sec:Aird_recovery}

\begin{figure*}[htbp]
    \centering
    \includegraphics[width=1\linewidth]{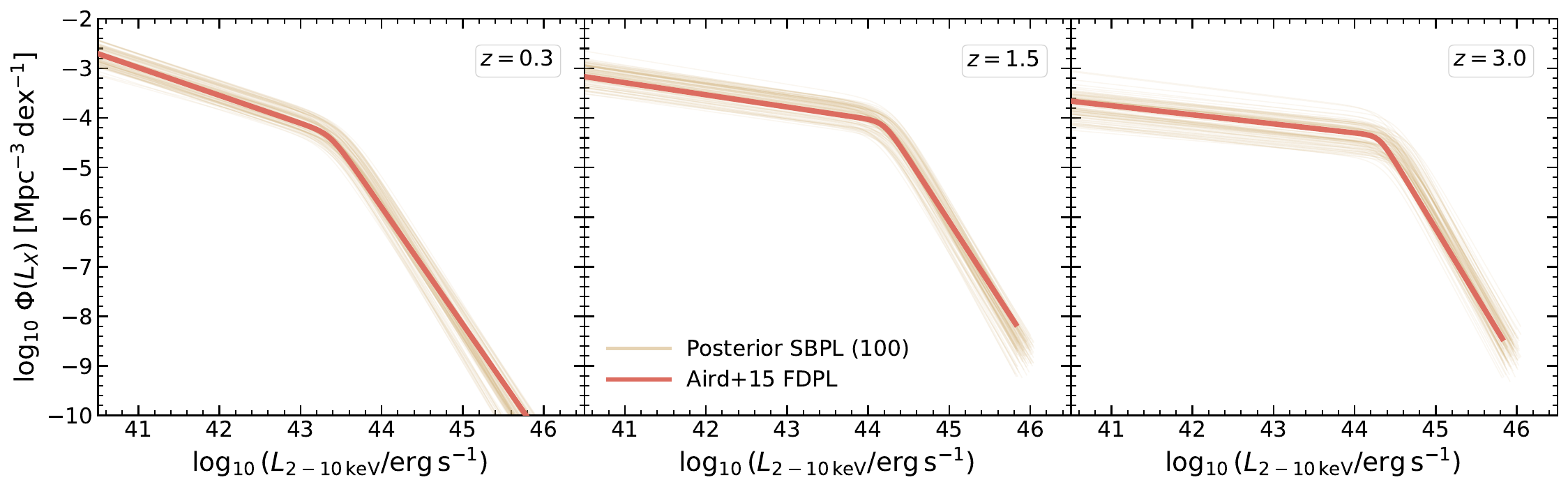}
    \caption{Mock-recovery test of the SBPL fitting framework (see \cref{sec:SBPL}) at three representative redshifts. Grey curves show 100 recovered SBPL posteriors, while the red curve indicates the input model used to generate the mock catalogue. The input recovery demonstrates that the fitting procedure can reproduce the underlying luminosity-function shape over the luminosity range probed by the mock data.}
    \label{fig:LDDE_recover}
\end{figure*}

To validate the SBPL approach, we test it using an AGN mock sample following the soft-band-selected FDPL model of \cite{Aird_2015}, providing a consistency check under assumptions matched as closely as possible to the present analysis. It does not estimate absorption-dependent completeness relative to the intrinsic AGN population. Such a calculation would require forward modelling the joint distribution $\Phi(L_{\rm X}, z, N_{\rm H})$, rather than a marginal XLF alone. \Cref{fig:LDDE_recover} shows the input FDPL model in comparison to recovered SBPL posteriors.



\section{Integrated BHAD comparison}
\label{sec:BHAD_app}

\begin{figure*}[htbp]
    \centering
    \includegraphics[width=1\linewidth]{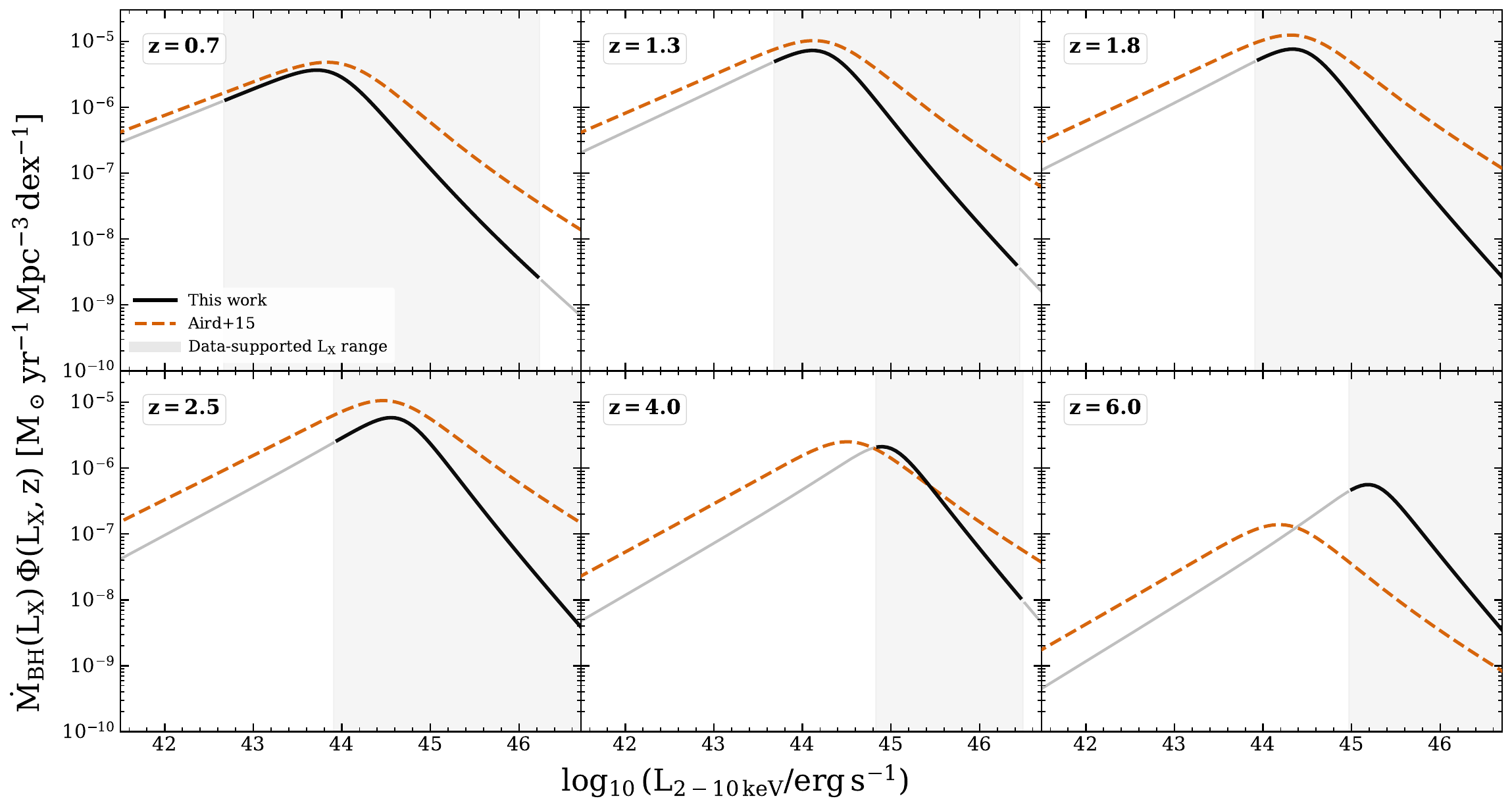}
    \caption{Luminosity-resolved contribution to the BH accretion-rate density at selected redshifts. Each panel shows the BHAD integrand, $\dot{M}_{\rm{BH}}(L_{X})\Phi(L_{X},z)$, as a function of rest-frame $2-10$ keV luminosity. Black curves show the best-fitting XLF from this work converted to accretion-rate density assuming $\xi=0.1$, while brown dashed curves show the corresponding \cite{Aird_2015} FDPL prediction. Grey shaded regions mark the luminosity range directly supported by the eROSITA data at each redshift. The figure illustrates which luminosities dominate SMBH growth at different epochs and highlights where differences between this work and \cite{Aird_2015} contribute to the offset in \cref{fig:BHAD}.}
    \label{fig:BHAD_app_pic}
\end{figure*}

To unpack differences in the observed BHADs of different XLF models, the mass accretion integrated XLF, \cref{fig:BHAD_app_pic}, illustrates offsets in a bolometric correction independent manner.

\end{appendix}
\end{document}